\documentclass[10pt,letterpaper,twocolumn,aps,pra,showpacs,superscriptaddress]{revtex4-1}
\pdfoutput=1

\usepackage{layouts}
\usepackage{physics}
\usepackage{ucs}
\usepackage{amsmath}
\usepackage{amsfonts}
\usepackage{amssymb}
\usepackage{makeidx}
\usepackage{cellspace,booktabs}
\usepackage{natbib}
\usepackage{hyperref}
\usepackage[capitalise]{cleveref} 
\usepackage{lipsum}
\usepackage{xcolor}
\usepackage{microtype}
\usepackage{bm}
\usepackage{graphicx}

\usepackage[binary-units]{siunitx}
\DeclareSIUnit{\dbm}{dBm}
\DeclareSIUnit{\photons}{photons}

\usepackage{8qbroNotation}

\definecolor{light-gray}{gray}{0.55}

\newcommand{\exv}[1]{ \left\langle #1 \right\rangle }

\begin{document}
\begin{abstract}
The duration and fidelity of qubit readout is a critical factor for applications in quantum information processing as it limits the fidelity of algorithms which reuse qubits after measurement or apply feedback based on the measurement result. Here we present fast multiplexed readout of five qubits in a single 1.2~GHz wide readout channel. Using a readout pulse length of 80~ns and populating readout resonators for less than 250~ns we find an average correct assignment probability for the five measured qubits to be $97\%$.
The differences between the individual readout errors and those found when measuring the qubits simultaneously are within $1\%$. We employ individual Purcell filters for each readout resonator to suppress off-resonant driving, which we characterize by the dephasing imposed on unintentionally measured qubits. We expect the here presented readout scheme to become particularly useful for the selective readout of individual qubits in multi-qubit quantum processors.
\end{abstract}

\date{\today}

\author{Johannes~Heinsoo}
\affiliation{Department of Physics, ETH Zurich, CH-8093 Zurich, Switzerland}
\author{Christian~Kraglund~Andersen}
\affiliation{Department of Physics, ETH Zurich, CH-8093 Zurich, Switzerland}
\author{Ants~Remm}
\affiliation{Department of Physics, ETH Zurich, CH-8093 Zurich, Switzerland}
\author{Sebastian~Krinner}
\affiliation{Department of Physics, ETH Zurich, CH-8093 Zurich, Switzerland}
\author{Theodore~Walter}
\affiliation{Department of Physics, ETH Zurich, CH-8093 Zurich, Switzerland}
\author{Yves~Salath\'e}
\affiliation{Department of Physics, ETH Zurich, CH-8093 Zurich, Switzerland}
\author{Simone~Gasperinetti}
\affiliation{Department of Physics, ETH Zurich, CH-8093 Zurich, Switzerland}
\author{Jean-Claude~Besse}
\affiliation{Department of Physics, ETH Zurich, CH-8093 Zurich, Switzerland}
\author{Anton~Poto\v{c}nik}
\affiliation{Department of Physics, ETH Zurich, CH-8093 Zurich, Switzerland}
\author{Christopher~Eichler}
\affiliation{Department of Physics, ETH Zurich, CH-8093 Zurich, Switzerland}
\author{Andreas~Wallraff}
\affiliation{Department of Physics, ETH Zurich, CH-8093 Zurich, Switzerland}

\title{Rapid high-fidelity multiplexed readout of superconducting qubits}

\maketitle

\section{Introduction}
An essential feature of any digital quantum computer or simulator is the ability to measure the state of multiple qubits with high fidelity. In particular, high-fidelity single-shot measurements are needed for determining the result of quantum computation~\cite{Nielsen1997}, observing error syndromes in quantum error correction~\cite{DiVincenzo2009,Barends2014} and for achieving high channel capacity in quantum communication protocols such as quantum teleportation~\cite{Bennett1993,Steffen2013}. Moreover, quantum non-demolition measurements are used for conditioning quantum state initialization~\cite{Johnson2012, Riste2012, Salathe2017}. Recent progress in scaling up quantum processors based on superconducting qubits has stimulated research towards multiplexed readout architectures with the goal of reducing device complexity and enhancing resource efficiency as discussed in more detail below~\cite{Jerger2012,Schmitt2014a,Jeffrey2014}.

Superconducting qubits are most commonly measured by employing their off-resonant coupling to a readout resonator~\cite{Blais2004, Wallraff2005}. This dispersive interaction results in a qubit-state dependent shift of the resonator frequency, which is probed using coherent microwave fields. Recent improvements in the efficiency of microwave parametric amplifiers~\cite{Caves1982, Yurke1996, Castellanos2008, Eichler2014a} have enabled single-shot dispersive qubit readout with high fidelity~\cite{Mallet2009, Vijay2011}. Furthermore, the use of Purcell filters~\cite{Reed2010, Jeffrey2014, Bronn2015b} led to the implementation of faster readout circuits resulting in a reduction of the readout time down to \SI{50}{\nano\second} for single qubits without introducing additional qubit decay~\cite{Walter2017}.

Extensions of dispersive readout to multiple qubits can be realized by either coupling multiple qubits to a single readout resonator~\cite{Filipp2009b,DiCarlo2010} or by probing several readout resonators coupled to a single feedline with a multi-frequency pulse~\cite{Jerger2012}. The latter approach allows for selective readout of any subset of the qubits by choosing the corresponding frequency components in the measurement pulse. High-fidelity frequency multiplexed readout has first been achieved with multiple bifurcation amplifiers~\cite{Schmitt2014a}, one for each qubit, and more recently by employing a single broadband parametric amplifier~\cite{Jeffrey2014, Mutus2014a, Macklin2015, Roy2015c}. Multiplexed readout with Purcell protection has been achieved by coupling multiple readout resonators to a single resonator based Purcell filter~\cite{Neill2017}. Broadband Purcell filters based on stepped impedance resonators have also been realized~\cite{Bronn2017a}. Other recent multi-qubit experiments either employ individual readout lines for each qubit~\cite{Corcoles2015, Reagor2017} or avoid Purcell decay of qubits by using narrowband readout resonators~\cite{Bultink2016, Asaad2016, Song2017, Neill2017, Bronn2017a}, which, however, increase the time required for high-fidelity qubit readout.

In this work, we demonstrate frequency multiplexed readout of up to five qubits using a single readout channel, see \cref{fig:concept}~(a) for a schematic of the concept. We use individual Purcell filters for each readout resonator, which in addition to protecting the qubits from Purcell decay, also suppress the off-resonant driving of untargeted readout resonators, thus avoiding the unintentional dephasing of qubits. We characterize this readout crosstalk in our experiments, by analyzing correlations in the readout between all pairs of qubits and by measuring the additional dephasing imposed on untargeted qubits during the readout.

The presented multiplexed readout concept is expected to be particularly useful in multi-qubit algorithms, in which subsets of qubits are measured while other qubits evolve coherently. In the surface code ~\cite{Fowler2012}, for example, a set of ancillary qubits is repeatedly measured while keeping all data qubits ideally unperturbed. Other examples for protocols relying on readout of individual qubits during the algorithm include the iterative quantum Fourier transform~\cite{Griffiths1996}, entanglement distillation~\cite{Bennett1996b}, and deterministic entanglement swapping~\cite{Yurke1992}.

\section{Concept of readout architecture}
\label{sec:concept}

\begin{figure}[t]
	\includegraphics[width=\linewidth]{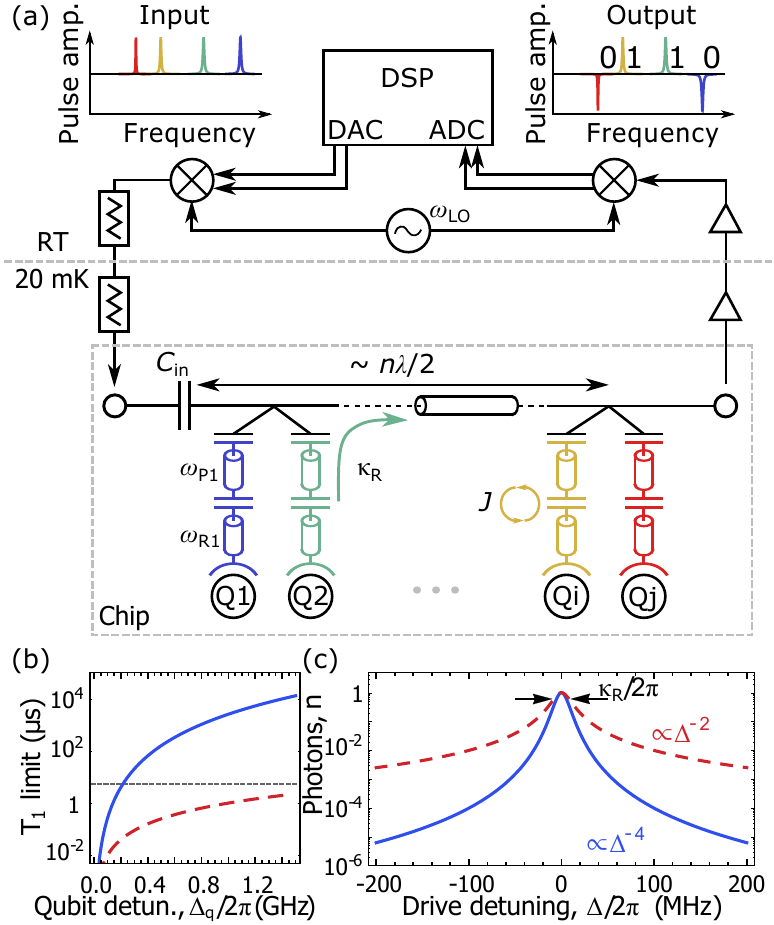}
	\caption{ \label{fig:concept}
(a) Schematic of the multiplexed readout experiment showing a circuit diagram of the superconducting chip at $T\approx \SI{20}{\milli\kelvin}$ and the room temperature (RT) electronics. Multi-frequency pulses used for readout are synthesized with a digital signal processing (DSP) unit, then upconverted to microwave frequencies by analog mixing with a local oscillator (LO) field, and applied to the input port of the feedline after several stages of attenuation. The readout signal emitted from the sample is amplified, downconverted and digitized with an analog-to-digital (ADC) converter and further processed with the same DSP unit used for pulse synthesis.
(b) Qubit lifetime \lifetime{} as limited by the Purcell effect \textit{vs.} qubit detuning with (blue solid line) and without (red dashed line) Purcell filter. The dashed line at $\lifetime{}=\SI{5}{\micro\second}$ indicates typical $T_1$ times measured in this work.
(c) Calculated photon number in the readout resonator with $\kappa_R/2\pi = \SI{20}{\mega\hertz}$ normalized to its maximum value as a function of drive detuning $\Delta = \freq{d} - \freq{\rr{}}$  with (blue solid line) and without (red dashed line) Purcell filter.}
\end{figure}

For readout we dispersively couple each qubit~\qb{i} to a resonator \rr{i} with resonance frequency $\freq{\rr{i}}$, see \cref{fig:concept}~(a). The readout resonator is coupled through a dedicated Purcell filter \pr{i} to a common feedline. The effective linewidth of the readout resonator is given by
\begin{equation}
\begin{split}
	\lwidth{\rr{}} &= \frac{1}{2} \left(\lwidth{\pr{}}-\Re{\sqrt{-16 \cplrrtopr{}^2+\left(\lwidth{\pr{}}-2 i \Delta_{\rr{},\pr{}}\right){}^2}}\right),
\end{split}
\end{equation}
with the linewidth of the Purcell filter~\lwidth{\pr{}}, the coupling strength~\cplrrtopr{} and detuning between readout resonator and Purcell filter $\Delta_{\rr{},\pr{}}=\freq{\rr{}}-\freq{\pr{}}$, see \cref{app:inputoutput} for details. In order to achieve fast readout we targeted an effective linewidth of $\lwidth{\rr{}}/2\pi \gtrsim \SI{10}{\mega\hertz}$. Taking a realistic detuning of $\Delta_{\rr{},\pr{}}/2\pi\times \lesssim \SI{5}{\mega\hertz}$ into account, which results from the finite accuracy of circuit design and fabrication, we design $\cplrrtopr{}/2\pi=\SI{10}{\mega\hertz}$ and $\lwidth{\pr{}}/2\pi=\SI{40}{\mega\hertz}$ to approach our targeted $\lwidth{\rr{}}$. Furthermore, the Purcell filter parameters are designed to strongly suppress qubit decay into the feedline ~\cite{Reed2010, Sete2015}. As illustrated in \cref{fig:concept}~(b), for typical detunings $\Delta_q=\omega_{\rm Q}-\omega_{\rm R}$ between qubit and resonator the $T_1$ limit imposed by Purcell decay through the readout resonator is expected to be significantly higher than the typical $T_1$ times measured in our current device.

For realizing frequency multiplexed readout all Purcell filters are coupled to a common feedline and have an approximately equal frequency spacing of $\Delta_{\rr{}}/2\pi\approx\SI{160}{\mega\hertz}$. Choosing this relatively small frequency spacing in combination with a large $\kappa_R$, could induce significant population in untargeted resonators while driving another resonator nearby in frequency. Such unintentional resonator population causes additional dephasing of untargeted qubits~\cite{Gambetta2006}. The use of dedicated Purcell filters, however, strongly suppresses the off-resonant driving of each individual readout resonator. In the limit of large drive detuning the intra-resonator photon number scales as $\propto \Delta^{-4}$ with a Purcell filter, as compared to $\propto \Delta^{-2}$ without it, see \cref{fig:concept}~(c).

\begin{figure*}[t]
\begin{center}
\includegraphics[width=\linewidth]{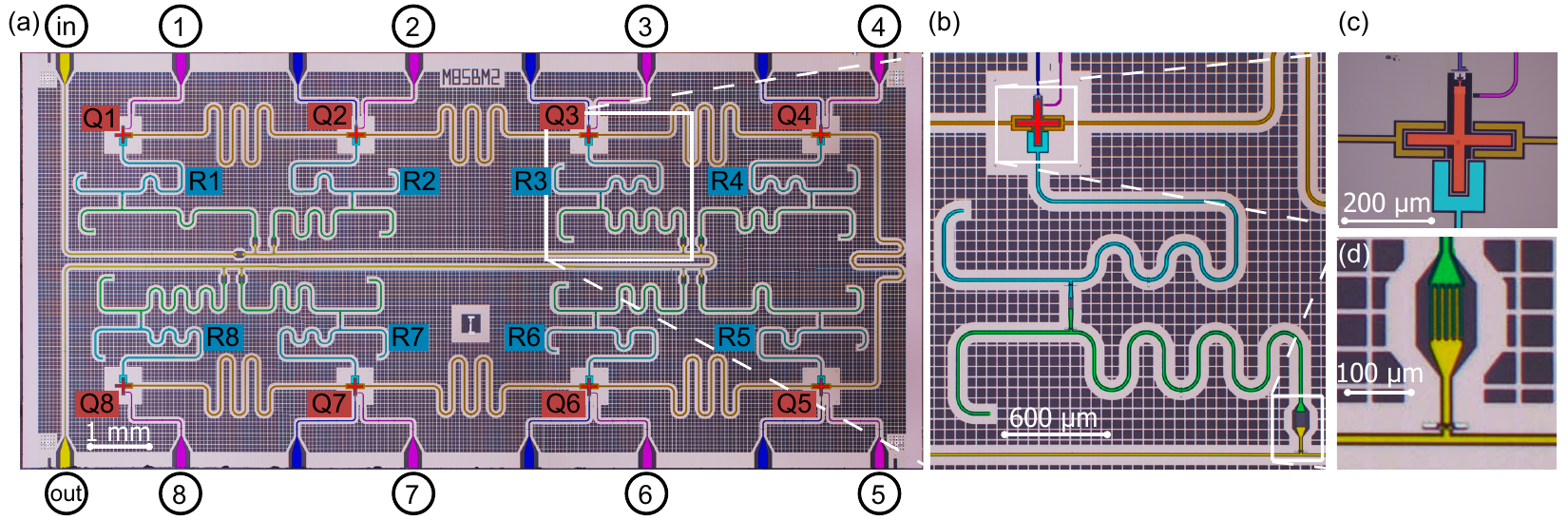}
\end{center}
\caption{
(a) False-colored optical micrograph of the device with qubits \qb{i} (red), readout resonators \rr{i} (blue), the Purcell filters (green), the coupling bus resonators (orange), the charge lines for single qubit manipulation (purple), the flux lines for single qubit tuning (dark blue) and the feedline (yellow). The ports used for probing the device are denoted in circles. (b) Enlarged view of \qb{3} with its readout resonator and Purcell filter. (c) Cross-shaped single island qubit capacitavely coupled to a readout resonator, qubit-qubit coupling resonators and a charge line, and inductively coupled to a flux-line. (d) Finger capacitor coupling the Purcell filter to the feedline.} \label{fig:device}
\end{figure*}

\newcommand{\pcomp}{\ensuremath{P_{\SI{1}{\decibel}}}}

To read out multiple qubits simultaneously we synthesize a multi-frequency probe pulse using a digital signal processing (DSP) unit and then upconvert, attenuate and apply the pulse to the input port of the feedline. A capacitor  $C_{\mathrm{in}}$ at the input provides directionality to the readout signal, which preferentially decays from the resonator towards the output port and thus minimizes signal loss into the input port. For the chosen capacitance of $C_{\mathrm{in}} = \SI{40}{\femto\farad}$, a proportion of $(1+\abs{\Gamma}^2)/2\approx \SI{98}{\percent}$ of the readout signal propagates towards the output port, where $\Gamma(\freq{}) = 1/(1 + 2i \freq{} Z_0 C_{\mathrm{in}})$ is the reflection coefficient of the capacitor, see \cref{app:inputoutput} for details. Moreover, the capacitor $C_{\mathrm{in}} $ enforces voltage antinodes at positions separated from it by integer multiples of half the wavelength~$n \wlen{\rr{i}}/2$, to which we couple the Purcell filters.

The output signal emitted from the sample is amplified by a traveling wave parametric amplifier (TWPA), a broad-band near-quantum-limited non-degenerate amplifier with an average gain of $\SI{20}{dB}$ in the relevant bandwidth $\SI{6.5}{}$--$\SI{7.8}{\giga\hertz}$ and a compression point of $\pcomp=\SI{-100}{\dbm}$~\cite{Macklin2015}. After several additional stages of amplification (see \cref{app:setup}) the readout signal is downconverted and digitized with the same DSP unit as used for pulse synthesis. As the DSP unit has a total bandwidth of \SI{1.2}{\giga\hertz} we can read out the state of up to eight qubits given our choice of detuning $\Delta_{\rr{}}$. The digitized signal is filtered in parallel for each readout frequency with a mode matched filter implemented by weighted integration. The combination of asymmetric feedline, a near quantum limited amplifier, and mode matched filtering results in a total average measurement efficiency of $\eta=\SI{49}{\percent}$, see \cref{app:efficiency}.

\section{Device description and characterization}
\label{sec:device}
We demonstrate the concept described above, with a device featuring eight single island transmon qubits~\cite{Koch2007, Barends2013}, see \cref{fig:device}.  Each qubit has an individual drive-line to perform single-qubit gates and all but Q1 and Q8 have a flux-line for frequency tuning,  as the number of ports on the sample mount is limited. While the readout resonators and Purcell filters are implemented as $\lambda/4$ resonators, qubit-qubit coupling resonators are realized as $\lambda/2$ resonators. The planar Nb and Al structures on the sapphire substrate of the device were defined using photo- and e-beam lithography, for fabrication details see~\cref{app:sampleparams}.

\begin{figure}
	\begin{center}
	\includegraphics[width=\linewidth]{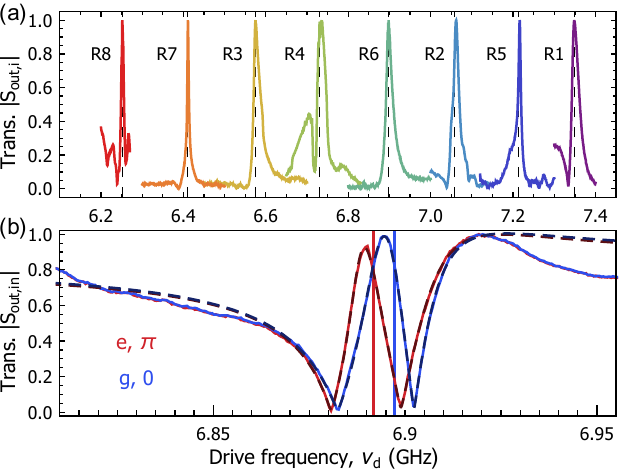}
	\end{center}
	\caption{
	(a) Transmission spectra $\abs{S_{\mathrm{out},i}(\nu_d)}$ of the readout resonators measured from the charge line ports $i$ to feedline output port as a function of drive frequency $\nu_d$. (b) Transmission spectrum  $\abs{S_{\mathrm{out},\mathrm{in}}(\nu_d)}$ measured from the feedline input to output port close to the resonance frequency of \rr{6} with (red line) and without (blue line) a $\pi$-pulse applied to qubit \qb{6} prior to measurement. The dashed lines are fits to the model described in \cref{app:inputoutput} with vertical lines indicating the fitted readout resonator frequencies for the two cases.
	}\label{fig:spectra}
\end{figure}

Transmission spectra measured from the eight qubit drive-lines to the output port reveal a single peak for each readout resonator, see \cref{fig:spectra} (a). The frequency spacing between individual resonator frequencies is close to the designed value of \SI{160}{\mega\hertz}. We extract linewidths $\lwidth{\rr{i}}/2\pi$ between~\SI{3}{\mega\hertz} and \SI{11}{\mega\hertz}. We attribute additional features in the measured spectra to the residual direct coupling between the drive-lines to other elements on the chip, as well as the finite detuning between the readout resonators and their corresponding Purcell filters.

We measure the state-dependent dispersive shift~\dshift{} for each qubit by preparing either the ground or excited state before probing the transmission from the input to the output port of the feedline, see \cref{fig:spectra}~(b) for example data for Q6. We observe a wide dip in the transmission spectrum resulting from the Purcell filter and an additional peak in the center close to the frequency of the readout resonator. The frequency of this peak depends on the qubit state while the background, dominated by the Purcell filter response, remains largely unaffected. The measured transmission data around a single resonance is well reproduced by the analytic expression obtained from the input--output theory, see \cref{app:inputoutput}. From fits of this model to all measured data sets we obtain the resonator parameters summarized in  \cref{tab:resonator}. The effective linewidths and dispersive shifts of most qubits are smaller than the target values discussed in \cref{sec:concept} owing to imprecisions in device fabrication. To achieve detunings $\Delta_{\rr{},\pr{}}$ between readout resonator and Purcell filter below $\SI{20}{\mega\hertz}$ across the entire sample we carefully modeled the microwave properties of the individual elements as discussed in~\cref{app:omegar}.

\begin{figure}
	\includegraphics[width=\linewidth]{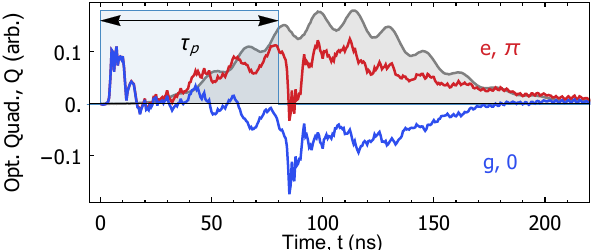}
	\caption{
The complex averaged time dependent response of resonator \rr{6} for \qb{6} prepared in the exited (red) and ground (blue) state in the quadrature with largest integrated difference. The light blue area indicates the applied readout pulse of length $\tau_p = \SI{80}{\nano\second}$ and the gray area marks the the difference between the response for the two states.
	}\label{fig:timetrace}
\end{figure}

We perform time-resolved measurements of the resonator response to a  \SI{80}{\nano\second} long probe pulse for both the qubit initially prepared in the ground and excited states, see \cref{fig:timetrace}. We show the measured response downconverted to the frequency of the probe pulse and chose the phase such that  the real part of the difference between the ground and excited state response is maximal. At the start and the end of the readout pulse we observe a peak and a dip, respectively, which are both independent of the qubit state. We attribute this feature to the fast ring-up and ring-down dynamics of the Purcell filter. In addition, we observe a smooth change in the difference between the ground and excited state responses stemming from the qubit-resonator dynamics. The oscillations in the signals and their difference result from a two-frequency beating caused by the finite quadrature imbalance of the downconversion mixer.

\begin{table}[b]
\centering
\begin{tabular*}{\linewidth}{l @{\extracolsep{\fill}} ccccc}
\hline\hline
 											& \rr{2}          & \rr{3}          & \rr{5}            & \rr{6}			& \rr{7}		\\ \hline
 $\freq{\rr{}}/2\pi$ (\SI{}{\giga\hertz})   & \num{7.058}     & \num{6.575}     & \num{7.214}     	& \num{6.898}		& \num{6.409} 	\\
 $\freq{\pr{}}/2\pi$ (\SI{}{\giga\hertz})   & \num{7.057}     & \num{6.580}     & \num{7.196}     	& \num{6.898}		& \num{6.392} 	\\
 $\lwidth{\pr{}}/2\pi$ (\SI{}{\mega\hertz}) & \num{32.2}      & \num{35.6}      & \num{57.8}      	& \num{38.3}		& \num{32.6}  	\\
 $\cplrrtopr{}/2\pi$ (\SI{}{\mega\hertz})	& \num{9.2}       & \num{7.9}       & \num{6.9}       	& \num{8.7}			& \num{7.8}   	\\
 $\lwidth{\rr{}}/2\pi$ (\SI{}{\mega\hertz}) & \num{14.3}      & \num{7.8}       & \num{4.5}			& \num{11.3}   		& \num{3.1}     \\
 $\dshift{}/2\pi$ (\SI{}{\mega\hertz})		& \num{-4.05}	  & \num{-1.11} 	& \num{-4.80} 		& \num{-2.66}		& \num{-1.92} 	\\
\hline\hline
\end{tabular*}
\caption{Parameters of readout resonator \rr{i} obtained from fits to transmission spectra equivalent to the one shown in \cref{fig:spectra} (b). The Purcell filter frequency~\freq{\pr{}}, readout resonator frequency~\freq{\rr{}}, Purcell filter linewidth~\lwidth{\pr{}} and their coupling rate~\cplrrtopr{}, the effective readout resonator linewidth~\lwidth{\rr{}} and dispersive shift~\dshift{} are listed.} \label{tab:resonator}
\end{table}

We measured the difference between ground and excited state response of all readout resonators, the complex conjugate of which we use as the integration weights in the DSP unit. Choosing this quantity as a mode matched filter is known to provide near optimal filter efficiency for a given readout frequency and power~\cite{Gambetta2007,Bultink2017}.

\section{Multiplexed single-shot readout}
\label{sec:single}

\begin{figure}[t]
\includegraphics[width=\linewidth]{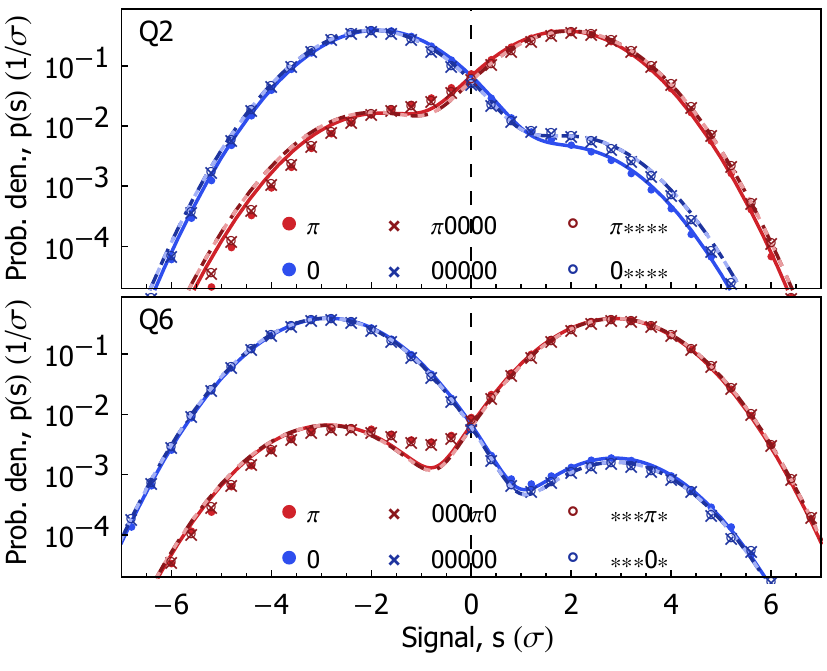}
\caption{Histograms of integrated single-shot readout signal of \qb{2} and \qb{6} respectively prepared with a $\pi$-pulse (red) and without (blue). The signal $s$ is normalized by the width of the gaussian distribution. The experiment data (dots), simultaneous fits to a double Gaussian distribution (solid lines) characterized with their different mean values and equal width. Individual readout, readout pulse applied also to all other qubits in a ground state (darker tones, $0$) or averaged over all states of other qubits (brighter tones, $*$).} \label{fig:histograms}
\end{figure}

We quantify the performance of single-shot readout for each qubit by preparing the qubit in either the ground or the excited state and by subsequently applying a readout pulse at the corresponding readout resonator frequency. The integrated response signal $s$ for the two input states follows a bimodal Gaussian distribution with the distribution width $\sigma$ as shown in \cref{fig:histograms} for the qubits \qb{2} and \qb{6} having the smallest and largest $\text{SNR} = (\exv{s}_\pi - \exv{s}_0)/\sigma$. We normalize $s$ by the width of the distribution $\sigma$ to make the SNR easily comparable for the different qubits. Each qubit state is prepared $\nrep \approx \num{1.3e6}$ times. In all experiments we also apply an additional readout pulse prior to the state preparation to herald the ground state~\cite{Johnson2012, Riste2012, Walter2017}. The heralding discards $\ptherm{}=~\SIrange{4}{6}{\percent}$ of the experiments for each qubit corresponding to the probability for the qubit to be thermally excited~\cite{Jin2015b}.

In order to assign a binary value corresponding to the outcome of the qubit measurement from the continuous valued signal $s$, we choose an assignment threshold, which best separates the prepared states of the qubit. We quantify the fidelity of the readout by the correct assignment probability $\pcor{}=[P(g|0)+P(e|\pi)]/2$, where $\pi$ ($0$) marks the state preparation with (without) a $\pi$-pulse, and $e$ ($g$) stands for the qubit assigned as in excited (ground) state. We maximize the assignment probability \pcor{} by optimizing the readout power and frequency for a given readout pulse length for each qubit individually.

The bimodal Gaussian fits to the single-shot histograms provide information on the sources of readout error~\cite{Walter2017}. There are three main error mechanisms: First, due to finite $\text{SNR}$, the two states cannot be fully distinguished because of the overlap of the two Gaussians. The overlap error accounts for less than $0.5\%$ error probability for qubits \qb{3,5,6\&{}7}. For \qb{2} this error amounts to $3.1\%$ owing to the lower readout power used for this qubit compared to all others. Qubit state mixing between the ground and excited states due to the readout tone~\cite{Boissonneault2009} causes an error probability $P(e|0)=\SIrange{0.1}{1}{\percent}$. Finally, when prepared in the excited state, the qubit may decay before or during the readout, which accounts for the reminder of the observed errors and ranges from $0.7\%$ for \qb{3} to $5.5\%$ for \qb{5} which has a combination of a slow readout resonator and low \lifetime{} compared to the other qubits, see \cref{app:sampleparams} for a comparison of parameters. Overall qubit decay appears to be the dominant source of error, which suggests that significant improvements in the readout performance are possible in future devices featuring longer $T_1$ times.

We repeat the single-shot readout experiment with probe pulses applied at all five readout frequencies simultaneously and with the qubits prepared in all $2^5 = 32$ combinations of basis states. From this dataset we first pick a subset, where all but one qubit are left in the ground state. The histograms with a single (dots) and multi qubit probe tone (crosses) are practically indistinguishable, see \cref{fig:histograms}. This indicates, that the probe tones do not have a significant spectral overlap with the mode matched filters of the other qubits. Moreover the signal distributions, obtained after averaging over all possible states of the other qubits (circles) are also almost identical. Thus each frequency component contains information about a single qubit only, which is confirmed by the nearly identical correct assignment probability for the individual readout \pcor{1} and 5-qubit readout $\pcor{5}$ shown in \cref{tab:fidelities}. The remaining discrepancy is on the level of variation of assignment probabilities in repeated experiments.

\begin{table}[b]
\centering
\begin{tabular*}{\linewidth}{l @{\extracolsep{\fill}} c c c c c}
\hline\hline \\
					& Q2				& Q3 				& Q5				& Q6				& Q7			\\ \hline
 \pcor{1}			& \num{94.2}\%		& \num{98.8}\%		& \num{93.6}\% 		& \num{97.9}\%		& \num{97.8}\%	\\
 \pcor{5}			& \num{94.5}\%		& \num{98.8}\%		& \num{92.9}\%		& \num{98.6}\%		& \num{97.9}\%	\\
\hline\hline
\end{tabular*}
\caption{Correct assignment probability \pcor{1} \& \pcor{5} of the single qubit \& simultanious 5 qubit single-shot readout correspondingly.}\label{tab:fidelities}
\end{table}

As the readout performance for each qubit remains largely undisturbed by the additional readout tones we use the individually obtained assignment threshold values and mode-matched filters. The ability to independently calibrate each subsystem is desired for system scalability.

\begin{figure}
\begin{center}
\includegraphics[width=\linewidth]{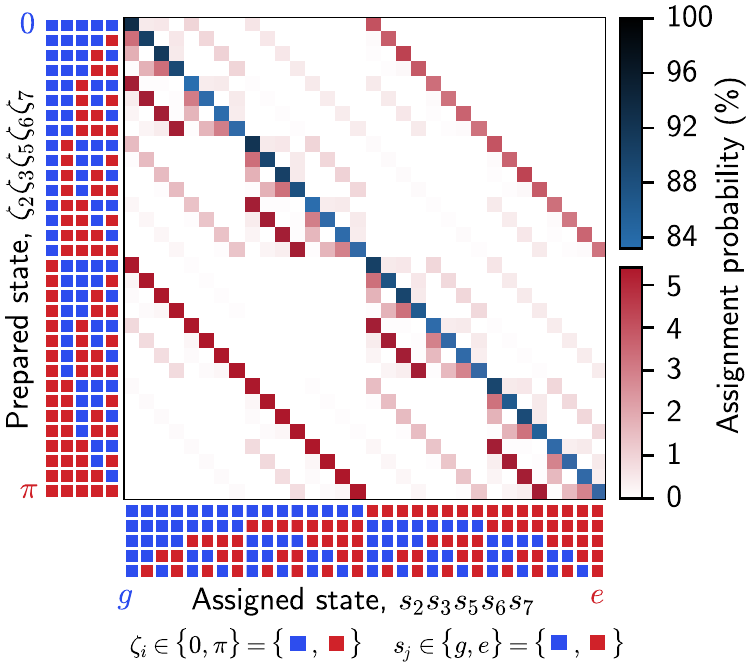}
\end{center}
\caption{Assignment probability matrix $P(s|\zeta)$ for each qubit \qb{i} prepared with a $\pi$-pulse $\zeta_i=\pi$ (red) or without a pulse $\zeta_i=0$ (blue) and each qubit assigned to either ground state $s_i=g$ (blue) or exited state $s_i=e$ (red). The qubits are ordered as \qb{2}, \qb{3}, \qb{5}, \qb{6} and \qb{7} from left to right (top to bottom) for the prepared (assigned) state.} \label{fig:probmatrix}
\end{figure}

The probability matrix $P(s_1\dotsm{}s_5|\zeta_1\dotsm{}\zeta_5)$ (\cref{fig:probmatrix}) of assigning state $s_i \in \left\lbrace e,g \right\rbrace$ for preparation $\zeta_i \in \left\lbrace 0,\pi \right\rbrace$ describes all state assignment. Ideally, $P(s|\zeta)$ is an identity matrix. The matrix obtained from the experimental data is close to diagonal with the largest deviation corresponding to assigning all qubits to the prepared excited states $P(eeeee|\pi\pi\pi\pi\pi)=\num{83.3}\,\%$ as this input state is most susceptible to individual qubit decay.
Apparent features in the full assignment probability matrix are the additional off-diagonal lines, which below the diagonal are indicative of individual qubit decay and above the diagonal of measurement induced excitation during the measurement. These off-diagonal elements are most pronounced for \qb{2} and \qb{5}, which have the largest decay and mixing errors. Moreover, as discussed in \cref{app:correlation}, the cross correlations extracted from the assignment probability matrix are up to $0.2\%$, which is small compared to the single qubit readout errors.

\section{Effect of measurement crosstalk on untargeted qubits}
\label{sec:crosstalk}

As the readout resonators are coupled to a common feedline and have a finite spectral overlap, the readout tone of qubit \qb{j} also populates the readout resonators of untargeted qubit \qb{i} with a qubit state dependent field amplitude $b_s$, which causes parasitic measurement induced dephasing \cite{Gambetta2006}. While the instantaneous measurement-induced dephasing rate $\dephrateinst(t) = 2 \dshift{} \Im{(b_g^{}(t) b_e^*(t))}$ changes during the measurement, the error per readout operation corresponds to the integrated effect of the probe pulse. Thus, we quantify the effect of measurement crosstalk on untargeted qubits as the average dephasing rate
$
	\dephrate_{ij} = 1/\measlen \int_{0}^{\infty} \dephrateinst_{ij}(t) \,dt
$
of \qb{i} due to the measurement of \qb{j} with a pulse length \measlen{}.

\newcommand{\rpamp}[1]{\ensuremath{\xi_{#1}}}

\begin{figure}
\begin{center}
\includegraphics[width=\linewidth]{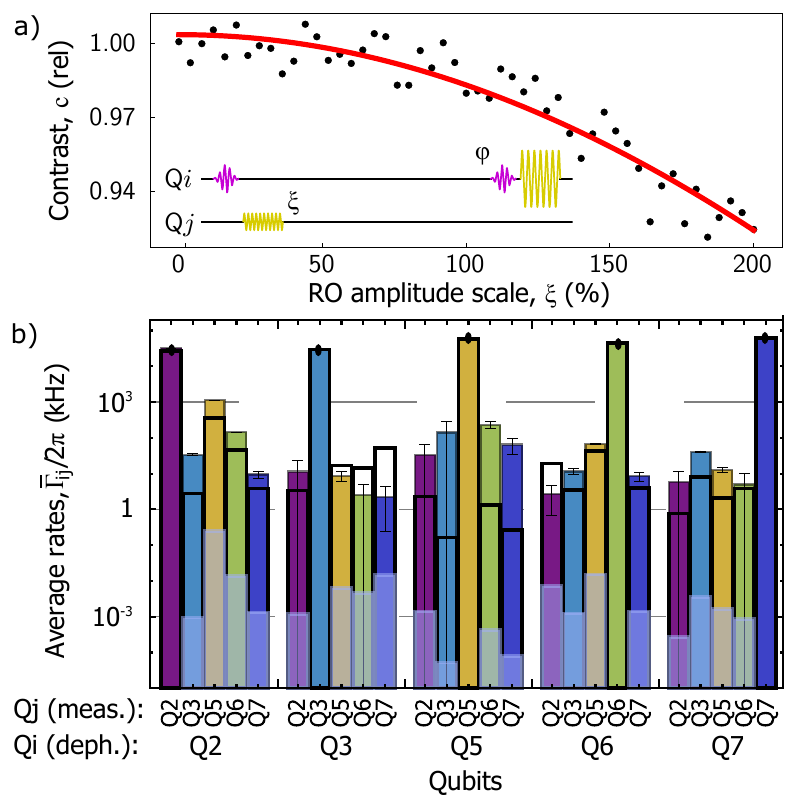}
\end{center}
\caption{(a) Contrast of the Ramsey oscillations $c$ of qubit \qb{7} as for different readout pulse amplitudes $\rpamp{}$ on \qb{3} (dots) and a Gaussian fit (red line). (a inset) Pulse scheme used for the dephasing measurement consisting of \SI{10}{\nano\second} $\pi$-pulses (pink) and \SI{80}{\nano\second} readout pulses (yellow).
(b) The average dephasing rate of qubit \qb{i} when applying a readout pulse to qubit \qb{j}. Experimentally measured rates (filled bars), calculated values based on the parameters extracted from spectroscopy shown in Table~\ref{tab:resonator} (thick black frames) and calculated dephasing rates for a Gaussian filtered probe pulse (transparent blue bars).
} \label{fig:dephasing}
\end{figure}

We measure the average dephasing rate in a Ramsey experiment~\cite{Slichter2012,Bultink2017} with the pulse scheme shown in the inset of \cref{fig:dephasing} (a).
By varying the phase $\phi$ of the second $\pi/2$-pulse on \qb{i} we observe Ramsey oscillations with a contrast $c$. In between the $\pi/2$-pulses we apply a probe pulse scaled in amplitude by a factor $\rpamp{}$ relative to the final probe pulse. As shown for the example data in \cref{fig:dephasing}~(a) for $i=7$ and $j=3$, the Ramsey contrast $c$ decreases with increasing $\rpamp{}$. We fit the observed data to the expression $c(\rpamp{}) = c_0 e^{-\dephrate \measlen \rpamp{}^2 }$ to extract the measurement induced dephasing rate $\dephrate{}$.  Here, the constant prefactor $c_0$ accounts for all other dephasing mechanisms which are independent of $\rpamp{}$.

When we apply the measurement pulse to the same qubit as the Ramsey experiment ($i=j$) we observe the intended measurement induced dephasing. As discussed in \cref{app:efficiency}, the measured $\dephrate_{ii}$ in combination with the SNR obtained from the histograms in \cref{fig:histograms} allows us to estimate the measurement efficiency $\eta$~\cite{Bultink2017}, which we find to be \SIrange{42}{52}{\percent}, mostly limited by the dissipative components before the TWPA and the internal loss of the TWPA.

The parasitic measurement induced dephasing $\dephrate_{ij}$ ($i \neq j$) spans two orders of magnitude (\cref{fig:dephasing}). For example the large dephasing of \qb{2} when measuring \qb{5} leads to a phase-error probability due to measurement-induced dephasing of $P_{\phi} = [1 - \text{exp}(- \dephrate_{ij} \measlen)]/2 \approx 11\%$ while for other qubit pairs the corresponding phase-error probability is below $0.1\,\%$.

To calculate the expected dephasing rate for the sample parameters given in \cref{app:sampleparams}, we solve for the field amplitude $b(t)$ in the readout resonator described by the equations of motion given in \cref{app:inputoutput}. The comparison of the calculated, depicted with black frame in \cref{fig:dephasing}, and the measured dephasing, depicted with filled bars, shows a qualitative agreement except for the dephasing of \qb{5}, for which we did not obtain reliable data due to qubit frequency instability and short dephasing time, see also Appendix~\ref{app:sampleparams}.

The good agreement with the model of parasitic measurement-induced dephasing justifies using the model for explaining the qualitative features and predicting possible future improvements. For example \qb{2} is most strongly dephased by the measurement tones in \rr{5} and \rr{6} as these are the readout resonators closest in frequency to \rr{2}. The readout pulse for \rr{5} dephases \qb{2} more strongly compared to the pulse for \rr{6}, since a much stronger tone was used for \rr{5} due to its small dispersive shift~\dshift{5}. In addition, \qb{2} shows the largest  measurement-induced dephasing in general, as it has the largest~\dshift{} and one of the largest~\lwidth{\rr{}} which leads to the largest spectral overlap with the probe pulses targeted to other readout resonators.

As the spectral overlap between the readout resonators is already small, the readout crosstalk is limited by the spectral width of the square shaped probe pulses, which are significantly wider in spectrum than the readout resonators. As shown in light blue in \cref{fig:dephasing}, by convolving the pulse shape with a Gaussian kernel with a width $\sigma=\SI{5}{\nano\second}$ the parasitic measurement-induced dephasing could possibly decrease by \numrange{2}{3} orders of magnitude. For the same $\lwidth{\rr{}}/\Delta_{\rr{}}$ ratio, but without individual Purcell filters such probe-pulse shaping results in only a minor improvement.

\section{Discussion and outlook}
\label{sec:outlook}
In this work we demonstrate frequency multiplexed readout of 5 qubits with high qubit selectivity. We show that the presented architecture enables fast readout in combination with low crosstalk. In particular, we show that the single qubit readout performance remains unaffected by the presence of multiple readout tones at a level where the individual readout calibration can be used for multi-qubit readout without degrading performance.

The primary source of errors in the single-shot qubit readout are single qubit decay and measurement-induced mixing. Furthermore, we found that the main source of readout crosstalk arose from the probe pulses spanning to the resonance of untargeted readout resonators. From simulations we expect that a significant reduction of parasitic resonator population could be achieved by a Gaussian filtering of the pulses.

Due to the short readout resonator occupation time of \SI{250}{\nano\second} and potentially low crosstalk, the readout architecture presented in this work seems particularly interesting for quantum error correction algorithms, in which a set of ancilla qubits is repeatedly measured while preserving the coherence of data qubits on the same chip~\cite{Versluis2017}.

\section*{Acknowledgments}
The authors would like to thank Philipp Kurpiers, Paul Magnard, Simon Storz, Adrian Beckert and Jacob Koenig for contributions to the experimental setup and Stefania Balasiu for contributions to the measurement control and analysis software. Moreover, the authors thank William D. Oliver for providing the TWPA.

The authors acknowledge financial support by the Office of the Director of National Intelligence (ODNI), Intelligence Advanced Research Projects Activity (IARPA), via the U.S. Army Research Office grant W911NF-16-1-0071, by the National Centre of Competence in Research Quantum Science and Technology (NCCR QSIT), a research instrument of the Swiss National Science Foundation (SNSF) and by ETH Zurich. The views and conclusions contained herein are those of the authors and should not be interpreted as necessarily representing the official policies or endorsements, either expressed or implied, of the ODNI, IARPA, or the U.S. Government.

\appendix
\section{Experimental setup}
\label{app:setup}
\begin{figure*}[bth]
\begin{center}
\includegraphics[width=\linewidth]{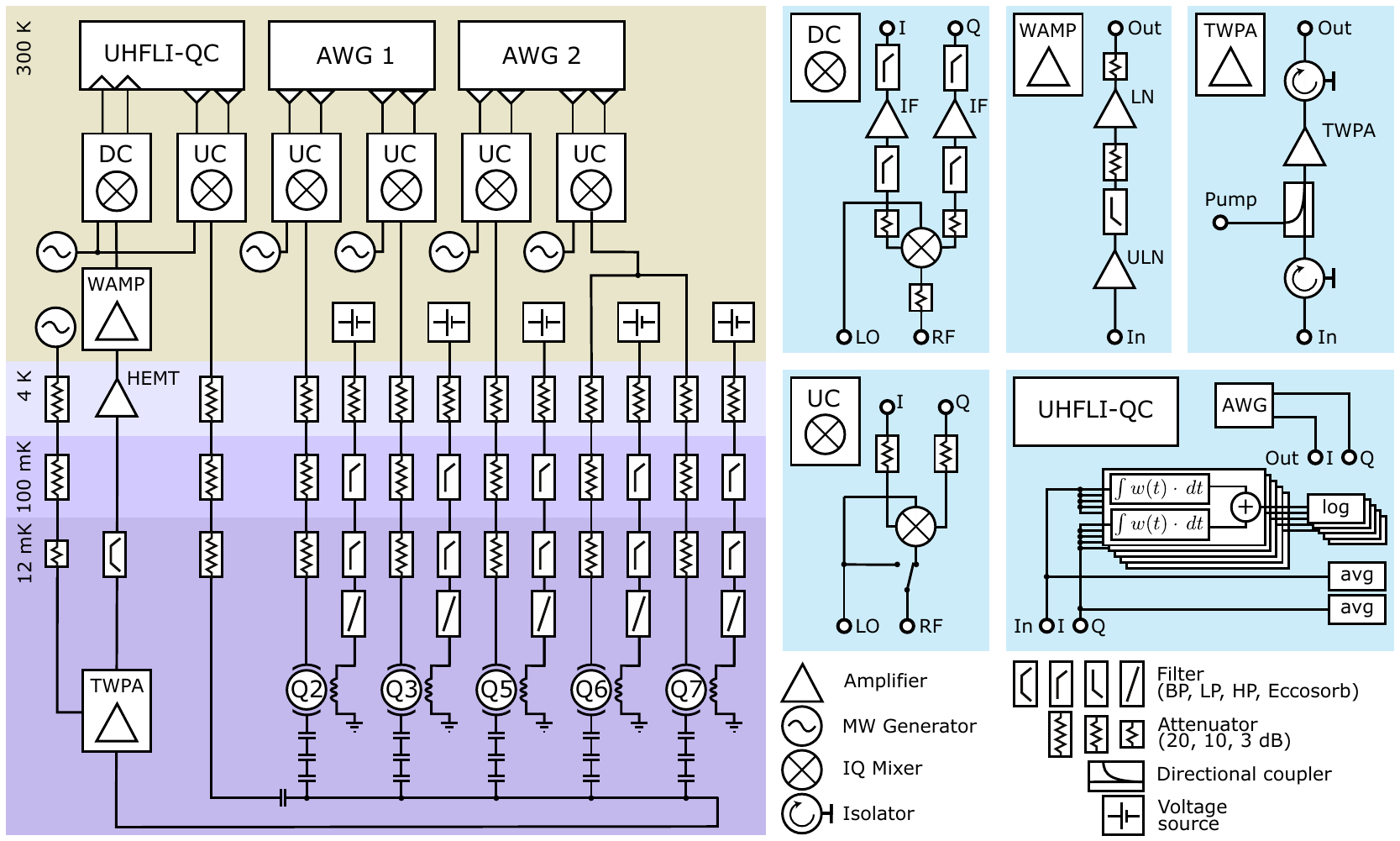}
\end{center}
\caption{Description of the electrical setup of the experiment.
The colored background on the left side indicates the different temperature stages for the components.
In the panels on the right we describe in more detail downconversion (DC), upconversion (UP), warm amplification (WAMP) boards, components used around traveling waveguide parametric amplifier (TWPA) and virtual hardware components of the Zurich Instruments UHFLI-QC. UHFLI-QC includes an internal arbitrary waveform generator (AWG), logging units (log), timetrace averager (AVG) and complex weighted integration units. Amplification chain includes a high electron mobility transitor (HEMT), intermediate frequency (IF), low noise (LN) and ultra low noise (ULN) amplifiers.} \label{fig:setup}
\end{figure*}

\newcommand{\namecompany}[1]{\textit{#1}}
\newcommand{\nameproduct}[1]{#1}
A multiplexed readout experiment requires instruments for generating probe pulses, detecting response signal and for manipulating the qubit states. The components for control are operated at different temperature stages of a cryogenic setup as shown in \cref{fig:setup}.

The readout pulses are generated and detected at \SI{1.8}{GSPS} using a single ultra-high frequency lock-in amplifier (\namecompany{Zurich Instruments} \nameproduct{UHFLI}). Upon receiving the readout trigger, the virtual AWG component of the built-in FPGA outputs a probe-pulse modulation waveform on the output channel pair. The two channels drive an IQ-mixer on an upconversion board, used for single-sideband upconversion to the radio frequency (RF). Along the way to the superconducting device, the probe pulses are attenuated by \SI{20}{\decibel} at the temperatures of \SI{4}{\kelvin}, \SI{100}{\milli\kelvin} and \SI{12}{\milli\kelvin} in an ${}^{3}\mathrm{He}$--${}^{4}\mathrm{He}$ dilution cryostat.

After the interaction with the readout resonators, the weak measurement signal is amplified using a wide bandwidth traveling waveguide parametric amplifier (TWPA)~\cite{Macklin2015} with the average gain of $G_{\mathrm{TWPA}}=\SI{20}{\decibel}$ and a 1dB-compression point of $\pcomp=\SI{-100}{\dbm}$. In order to impedance match the TWPA with its environment, the TWPA is surrounded with a wide-bandwidth isolator on both sides. The pump tone for the TWPA is generated at room-temperature and is added to the signal using a directional coupler at the TWPA input. Before the response pulse arrives at the room-temperature electronics it is bandpass-filtered at the base temperature and further amplified with a high-electron-mobility transistor (HEMT) amplifier at the \SI{4}{\kelvin} stage.

At room temperature the response pulse is amplified with an ultra low noise (ULN) and a low noise (LN) amplifier. The RF amplification is followed by downconversion to an intermediate frequency (IF) and the IF signal is subsequently further amplified. The warm amplification (WAMP) and downconversion (DC) boards feature additional filters and attenuation to suppress out-of-band noise and standing waves. A single local oscillator (LO) is shared by the measurement tone up- and downconversion.

The response signal is digitized by the UHFLI and passed through different digital signal processing components in the FPGA. The first FPGA component averages the incoming timetraces over all trigger events and is used for the data shown in \cref{fig:timetrace}. The single-shot measurement results are collected using parallel weighted integration and logging units. The on-board real-time data analysis gives a significant reduction of data and reduces the time used for uploading and analyzing the high-bandwidth data on a PC.

Each qubit is capacitively coupled to a drive line and inductively to a flux line. The drive pulses are generated using a channel pair of an AWG at \SI{1.2}{GSPS}. The single-sideband upconversion board has a built-in switch which allows bypassing the mixer without recabling for spectroscopy measurements. The qubit drive lines have the same cold attenuation configuration as the probe tone input. For parking the qubits in the frequency space, a voltage bias is directly applied to the flux-lines. At the \SI{100}{\milli\kelvin} and \SI{12}{\milli\kelvin} temperature stages the flux lines are not attenuated, but low-pass filtered with a cutoff at \SI{1}{\giga\hertz}. In addition, the flux lines have \nameproduct{EccoSorb} filters at the base temperature to suppress high frequency signals left unattenuated by the conventional lowpass filters.

\section{Detailed sample description}
\label{app:sampleparams}

\begin{table*}[t]
\centering
\begin{tabular*}{\linewidth}{l @{\extracolsep{\fill}} c c c c c}
\hline\hline
                                               & Q2              & Q3              & Q5              & Q6              & Q7          \\ \hline
Qubit frequency, $\omega_{\text{Q}}/2\pi$ (\SI{}{\giga\hertz})	& \num{6.254}     & \num{5.206}     & \num{6.441} (\num{5.457})     & \num{5.902}     & \num{5.442} \\
Max. qubit frequency, $\omega_{\text{Q,max}}/2\pi$ (\SI{}{\giga\hertz})                    & \num{6.260}     & \num{5.216}     & \num{6.996}     & \num{5.996}     & \num{5.442} \\
Qubit anharmonicity,  $\alpha/2\pi$ (\SI{}{\mega\hertz})	& \num{-226}       & \num{-246}       & \num{-198}       & \num{-234}       & \num{-238}   \\
Qubit lifetime, $T_1$ ($\mu$s) & \num{5.7} & \num{6.0} & \num{4.9} & \num{5.8} & \num{5.8} \\
Qubit coherence time, $T_2^*$ (\SI{}{\micro\second})                              & \num{4.1} & \num{2.5} & \num{0.7} & \num{3.1} & \num{7.8} \\
Thermal population, \ptherm (\%)              				       & \num{4.7}       & \num{5.1}       & \num{2.9}       & \num{6.0}       & \num{6.4}    \\
 \\
Readout resonator frequency, $\freq{\rr{}}/2\pi$ (\SI{}{\giga\hertz})   & \num{7.058}     & \num{6.575}     & \num{7.214} (\num{7.200})     	& \num{6.898}		& \num{6.409} 	\\
Purcell filter frequency, $\freq{\pr{}}/2\pi$ (\SI{}{\giga\hertz})   & \num{7.057}     & \num{6.580}     & \num{7.196}     	& \num{6.898}		& \num{6.392} 	\\
Purcell filter linewidth, $\lwidth{\pr{}}/2\pi$ (\SI{}{\mega\hertz}) & \num{32.2}      & \num{35.6}      & \num{57.8}      	& \num{38.3}		& \num{32.6}  	\\
Readout-Purcell coupling, $\cplrrtopr{}/2\pi$ (\SI{}{\mega\hertz})	& \num{9.2}       & \num{7.9}       & \num{6.9}       	& \num{8.7}			& \num{7.8}   	\\
Effective readout linewidth, $\lwidth{\rr{}}/2\pi$ (\SI{}{\mega\hertz}) & \num{14.3}      & \num{7.8}       & \num{4.5} (\num{11.8})		& \num{11.3}   		& \num{3.1}     \\
Qubit-coupling to \rr{i}, $g/2\pi$ (\SI{}{\mega\hertz})                                & \num{122.3}     & \num{123.4}     & \num{134.0}     & \num{115.9}     & \num{108.2} \\
Dispersive shift, $\chi/2\pi$ (\SI{}{\mega\hertz})                             & \num{-4.1}      & \num{-1.7}      & \num{-4.8} (\num{-0.9})      & \num{-2.6}      & \num{-2.4}   \\
Readout frequency, $\omega_{\text{RO}}/2\pi$ (\SI{}{\giga\hertz})                       & \num{7.056}     & \num{6.572}     & \num{7.208}     & \num{6.891}     & \num{6.407} \\
Readout IF, $\nu_{\text{RO mod.}}$ (\SI{}{\mega\hertz})                   & \num{195}       & \num{-289}      & \num{347}       & \num{30}        & \num{-454}   \\
\\
Readout photons, $n_{\text{RO}}$ & \num{4.1}       & \num{22.2}        & \num{2.9} (\num{126})       & \num{5.8}        & \num{9.7}   \\
Critical photons, $n_{\text{crit}}$ & \num{10.8}       & \num{30.2}        & \num{8.2} (\num{42})       & \num{18.2}        & \num{19.9}   \\
\hline\hline
\end{tabular*}
\caption{
Overview of qubit and resonator properties, see Appendix~\ref{app:sampleparams} for details. The numbers in parentheses for \qb{5} are the settings used for the dephasing measurements in Sec.~\ref{sec:crosstalk}.
} \label{tab:allparams}
\end{table*}

In \cref{tab:allparams} we list a detailed overview of the qubit parameters used in this work. Qubit \qb{5} exhibited from significant frequency instability and thus appeared at different configurations for the measurement discussed in \cref{sec:crosstalk} shown in parenthesis in Table~\ref{tab:allparams}. The maximum qubit frequency $\nu_{\text{Q,max}}$ is extracted from a two-parameter sweep of the voltage bias and the readout drive frequency to find the sweetspot of the qubit followed by a Ramsey experiment to extract the precise qubit frequency. The anharmonicity of each qubit is extracted by observing two-photon transitions in high-power spectroscopy. The qubit energy relaxation time $T_1$ and Ramsey decoherence time $T_2^*$ are characterized by standard timedomain experiments. The thermal population of the excited state $P_{\text{therm.}}$ is the probability that the qubit was found to be in the excited state in the preselection readout conducted before every single shot experiment run. For all qubits we used a $\pi$-pulse length of \SI{50}{\nano\s}.

The parameters related to the readout resonators and Purcell filters are obtained as explained in the main text. The frequencies $\nu_{\text{RO mod.}}$ are the intermediate frequencies of the probe pulse and and $\omega_{\text{RO}}/2\pi$ are the corresponding frequencies of the upconverted pulse. The photon number during the readout, $n_{RO}$, is measured using an AC-stark shift measurement and the critical photon number $n_{crit} = g^2/[4(\omega_Q - \omega_R)^2]$ is calculated from parameters above.

The device was fabricated on a substrate of c-plane cut single side polished sapphire from Rubicon Technology. A \SI{150}{\nano\meter} thin niobium film was deposited by StarCryoelectronics on wafers cleaned in ultrasound by \SI{50}{C} acetone and IPA. The rest of the circuit, except the Josephson junctions, is defined by optical lithography and dry etching process. Josephson junctions are formed by Al/AlOx/Al deposited in an electron-beam Plassys evaporator with a Dolan bridge shadow evaporation technique. Native Nb oxide was removed using ion milling before and after defining the e-beam mask.

\section{Input--output theory}
\label{app:inputoutput}
As discussed in the main text, the input port of the sample is interrupted by a capacitor in order to improve the efficiency of the qubit state measurement. This impedance mismatch in the feedline, however, changes the density of states in the feedline and therefore the effective linewidth of both the readout resonators and the Purcell filters. In order to understand and predict both the transmission spectrum and readout resonator time dynamics we construct here an input--output model of two coupled cavities $a$ and $b$ representing the Purcell filter and readout resonator respectively~\cite{Gardiner1985}.

The equation of motion for this two-mode system is given as
\begin{subequations}
\begin{align}
	\frac{da}{dt} &= -i\Delta_a a - \frac{\kappa_a + \gamma_a}{2} a -i J b + \sqrt{\kappa_a} a_i, \\
	\frac{db}{dt} &= -i\Delta_b b - \frac{\kappa_b + \gamma_b}{2} b -i J a + \sqrt{\kappa_b} b_i
\end{align}
\label{eq:resEq2M}%
\end{subequations}%
where $\Delta_{\{a,b\}}=\omega_{\{a,b\}} - \omega_d$ is the detuning between the drive frequency of the input field $\omega_d$ and the bare resonance frequency $\omega_{a,b}$ of the respective mode. The rates $\gamma_a$ and $\gamma_b$ are the internal loss rates of the resonators. The corresponding input--output relations are
\begin{subequations}
\begin{align}
	a_o &= a_i - \sqrt{\kappa_a} a, \\
	b_o &= b_i - \sqrt{\kappa_b} b,
\end{align}	
\label{eq:resCoupl2M}%
\end{subequations}%
where $\kappa_a$ is the large coupling of the Purcell filter to the feedline and $\kappa_b$ is the weak coupling of the readout resonator to the qubit drive-line.

A t-junction, such as the connection of the Purcell filter to the feedline, has three ports connected to the Purcell filter mode $a$, a port on the right $r$ and a port on the left $l$. For a symmetric, energy conserving and reciprocal 3-port device the relations between the incoming ($a_i$, $l_i$ and $r_i$) and outgoing ($a_o$, $l_o$ and $r_o$) mode-amplitudes are
\begin{alignat}{4}
	l_o &= -&&\frac{1}{3} l_i\, +\, &&\frac{2}{3} r_i \,+\, &&\frac{2}{3} a_o, \\
	r_o &=  &&\frac{2}{3} l_i\, -\, &&\frac{1}{3} r_i \,+\, &&\frac{2}{3} a_o, \\
	a_i &=  &&\frac{2}{3} l_i\, +\, &&\frac{2}{3} r_i \,-\, &&\frac{1}{3} a_o
\label{eq:tCoupl}
\end{alignat}
where we labeled the incoming and outgoing modes with respect to the Purcell filter.
At the input capacitor, a mode $c$ is connected with the port $l$ of the t-junction. The input--output relations of the capacitor in series are:
\begin{subequations}
\begin{align}
	c_o &= (1-\Gamma) l_o + \Gamma c_i \\	
	l_i &= (1-\Gamma) c_i + \Gamma l_o,
	\label{eq:boundM}%
\end{align}%
\end{subequations}%
where $\Gamma(\omega) = 1/(1 + 2i \omega Z_0 C_{in})$ with the frequency $\omega$, the characteristic impedance $Z_0$ and the input capacitor capacitance $C_{in}$.

Combining all the input--output relations and assuming a neglectable dispersion between the t-junction and the capacitor, we eliminate the modes $l$ and $a$ such that the equations of motion becomes
\begin{subequations}
\begin{align}
	\frac{da}{dt} &= -i\tilde{\Delta}_a a - \frac{\tilde{\kappa}_a + \gamma_a}{2} a -i J b + \frac{\sqrt{\kappa_a}}{2}\left(\tilde{c}_i+\tilde{r}_i\right),
	\label{eq:resEq2M2a} \\
	\frac{db}{dt} &= -i\Delta_b b - \frac{\kappa_b + \gamma_b}{2} b -i J a + \sqrt{\kappa_b} b_i \label{eq:resEq2M2b}%
\end{align}%
\label{eq:resEq2M2}%
\end{subequations}%
where $\Gamma$ alters the detuning from the cavity resonance $\tilde{\Delta}_a = \tilde{\omega}_a - \omega_d$, the coupling to the capacitor side input of the transmission line $\tilde{c}_i = (1-\Gamma)c$ and the coupling to the galvanically coupled port of the feedline $\tilde{r}_i = (1+\Gamma)r$. The linewidth, taking into account the altered environment, is
\begin{equation}
	\tilde{\kappa}_a = \kappa_a \frac{1+\Re{\Gamma}}{2}
	\label{eq:rescKappa}
\end{equation}
and similarly the resonator frequency $\tilde{\omega}$ becomes
\begin{equation}
	\tilde{\omega}_a = \omega_a + \kappa\frac{\Im{\Gamma}}{4}
	\label{eq:rescOmega}
\end{equation}
where $\kappa_a$ and $\omega_a$ correspond to cavity linewidth and frequency in the limit $\Gamma \rightarrow 1$ and the Purcell filter sees a single port. The input--output relations corresponding to the equations of motion above now read
\begin{subequations}
\begin{align}
		c_o &= c_i + (1-\Gamma) r_i - \frac{\sqrt{\kappa_a}}{2} (1-\Gamma) a, \\
		r_o &= r_i + (1-\Gamma) c_i - \frac{\sqrt{\kappa_a}}{2} (1+\Gamma) a, \\
		b_o &= b_i - \sqrt{\kappa_b} b.
\end{align}%
\label{eq:resCoupl2M2}%
\end{subequations}%

To extract the scattering parameters of the system, we solve Eq.~\ref{eq:resEq2M2} for the steady state and substitute the solution into Eq.~\ref{eq:resCoupl2M2}. By setting $r_i=b_i=0$ this approach yields the transmission coefficient through the feedline
	\begin{multline}
		\frac{S_{21}}{1-\Gamma} = \frac{r_o/c_i}{1-\Gamma} = 1 - \frac{1+\Gamma}{1+\Re{\Gamma}}  \\
		 \times\frac{ \tilde{\kappa}_a \left( \gamma_b + 2 i \Delta_b + \kappa_b\right) }{4 J^2 + \left(\gamma_a + 2 i \tilde{\Delta}_a +\tilde{\kappa}_a\right) \left(\gamma_b + 2 i \overset{ }{\Delta}_b + \kappa _b\right) },
		\label{eq:sparams21}
	\end{multline}
normalized by the insertion loss induced by the input coupler of the feedline. Equation~\eqref{eq:sparams21} is the model fitted to the data in \cref{fig:spectra}~(b). To obtain the transmission coefficient from the weekly coupled qubit drive line to the output port of the transmission line we set $r_i=c_i=0$:
	\begin{multline}
		S_{23} = \frac{r_o}{b_i} = \frac{1+\Gamma}{\sqrt{2\left(1+\Re{\Gamma}\right)}} \\
		\times\frac{4 i J \sqrt{\kappa _a} \sqrt{\kappa _b}}{4 J^2 + \left(\gamma_a + 2 i \tilde{\Delta}_a +\tilde{\kappa}_a\right) \left(\gamma_b + 2 i \overset{ }{\Delta}_b + \kappa _b\right) }
		\label{eq:sparams23}
	\end{multline}	
which describes the data shown in \cref{fig:spectra}~(a).

The inverse linewidth of readout resonator sets, as discussed in the main text, a limit to the readout time. When the Purcell filter decay rate $\kappa_a$ is significantly larger than the coupling rate $J$, the effective linewidth of the readout resonator becomes~\cite{Sete2015}
\begin{equation}
	\kappa_{\mathrm{R}}  = \frac{4 J^2 \tilde{\kappa}_a}{\tilde{\kappa}_a^2+4 \tilde{\Delta}_{\text{ab}}^2},
	\label{eq:kappaeffAprox}
\end{equation}
here neglecting all effects of $\gamma_{a,b}$ and $\kappa_b$ as these are typically small. However, we need to place $N$ readout structures into the finite detection bandwidth $\Delta_D$. Therefore, to avoid crosstalk from spectral overlap of the readout resonators, we have a practical limit to resonator linewidth $\kappa_a \lesssim \Delta_D/(n N)$, where  $n \sim 4$ is the amount of linewidths between each resonator frequency. Thus the assumption $\kappa_a \gg J$ may break down. To extract the exact expression for $\kappa_{\mathrm{R}}$ Eq.~\ref{eq:resEq2M2} is diagonalized and the real part of the eigenvalue corresponding to the readout resonator mode is the effective linewidth. Thus, we obtain
\begin{align}
	\kappa_{\mathrm{R}} = \frac{1}{2} \left(\tilde{\kappa}_a-\Re{\sqrt{-16 J^2+\left(\tilde{\kappa}_a-2 i \tilde{\Delta}_{\text{ab}}\right){}^2}}\right),
\end{align}
which is the expression used to calculate the readout resonator linewidth in Table~\ref{tab:resonator} from the fitted parameters. In the main text we denote the Purcell resonator frequency $\tilde{\omega}_a=\freq{\pr{}}$ and the linewidth $\tilde{\kappa}_a = \lwidth{\pr{}}$.

\section{Calculating the resonator frequencies}
\label{app:omegar}
Both the Purcell filter and the readout resonator for each qubit are realized as $\lambda / 4$-resonators on the device used in this work. The open ends of the $\lambda / 4$-resonators are capacitively coupled to either a qubit or to the feedline, see \cref{fig:device}~(c) and (d). Moreover, each pair of Purcell filter and readout resonator are coupled together with a capacitor, $C_c$, at a position $x_c$ from the terminated end of the resonators. The field amplitude between the terminated end and the coupling point $x=x_c$ and between the coupling point and the open end is described by the standard wave equation. Thus, the mode function for the phase variable $\phi(x,t)$ (time-integral of the voltage) of each resonator is given as~\cite{Bourassa2012}
\begin{align}
	\phi(x,t) = \phi_0(t) \times \begin{cases} B \sin(k x) & \text{ for } 0 \leq x \leq x_c \\
	\cos[k(x-d) - \theta] & \text{ for } x_c \leq x \leq d,\end{cases}
\end{align}
where $\phi_0$ is the time dependent field amplitude, $d$ is the length of the resonator, $B$ is a unitless scaling factor set by the boundary condition, $\theta$ is a phase offset, $k = \omega / v$ is the wave number and $\omega$ is the resonance frequency of the mode. Moreover, $v = 1/\sqrt{lc}$ is the phase velocity of field with $l$ and $c$ as the inductance and capacitance per length of the resonator respectively. This choice of mode function explicitly sets the boundary condition at $x=0$ where the resonator is grounded, while the phase $\theta$ is set by the boundary condition at the position $d$. As shown in~\cite{Bourassa2012}, the Euler-Lagrange equation at $d$ gives the equations of motion
\begin{align}
	\ddot{\phi}(d,t) + \frac{1}{C_0 l} \frac{\partial}{\partial x} \phi(x,t)\bigg\vert_{x=d} = 0,
	\label{eq:eqmoteulerlagrange}
\end{align}
where $C_0$ is the sum of capacitance to ground and to the qubit (feedline) for the readout resonators (Purcell filters). From the wave equation it follows that $\ddot{\phi}(x,t) = -\omega^2 \phi(x,t)$, such that Eq.~\eqref{eq:eqmoteulerlagrange} can be re-written into
\begin{align}
\tan (\theta) = C_0 Z_0 \omega \label{eq:d}
\end{align}
where $Z_0 = \sqrt{l/c}$ is the characteristic impedance of the co-planar waveguide. The final boundary condition to consider is at the coupling capacitor at $x=x_c$. Since the mode has to be continuous the scaling factor becomes
\begin{align}
B = \frac{\cos[k(x_c - d) - \theta]}{\sin(k x_c)}.
\end{align}
As the sum of the currents at $x_c$ has to be zero, we get the corresponding equation of motion~\cite{Bourassa2012}
\begin{align}
\ddot{\phi}(x_c,t) + \frac{1}{C_c l} \Bigg( \frac{\partial}{\partial x} \phi(x,t) \bigg\vert_{x=x_{c+}} - \frac{\partial}{\partial x} \phi(x,t) \bigg\vert_{x=x_{c-}} \Bigg) = 0
\end{align}
where $C_c$ is the coupling capacitance. As above this equation reduces to
\begin{align}
\tan(k x_c)^{-1} + \tan[k (x_c - d) - \theta] = C_c Z_0 \omega. \label{eq:xc}
\end{align}
Now we can solve Eqs.~\eqref{eq:d} and \eqref{eq:xc} numerically to find $\omega$ and $\theta$ for a given resonator length and with the capacitances $C_0$ and $C_c$ obtained from finite element simulations. Using these solutions we can accurately predict the frequency of each resonator, which ensures that the readout resonators are on resonance with its Purcell filters and that we achieve equidistant spacing of the resonators of the different qubits.

\section{Correlations and cross assignment fidelity output}
\label{app:correlation}
There are in general many ways of quantifying crosstalk in the state assignment for different qubits and here we will discuss cross fidelity and cross correlations.

\begin{figure}
\begin{center}
\includegraphics[width=\linewidth]{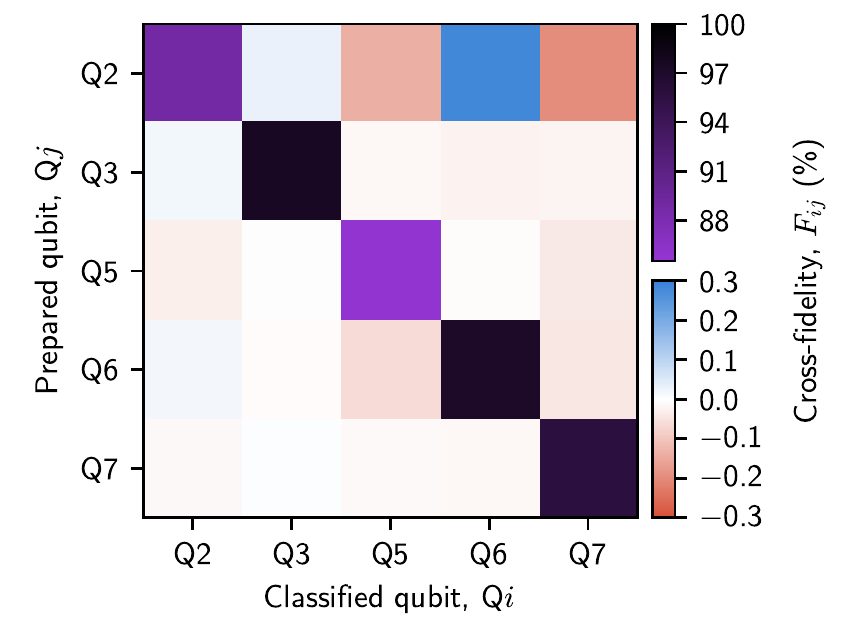}
\end{center}
\caption{Cross-fidelity calculated using Eq.~\eqref{eq:crossfid}. The diagonal elements correspond to the single qubit readout fidelities. } \label{fig:crossfid}
\end{figure}

In the main text we present the probabilities for assigning each multi-qubit state to the prepared state, which we denote $P(s | \zeta)$ with $s$ referring to the assigned state and $\zeta$ to the prepared state. From the marginal distributions, we can now quantify the information of qubit \qb{j} in the assignment of qubit \qb{i} by the cross-fidelity:
\begin{align}
F_{ij} = \exv{1 - P(e_i | 0_j) - P(g_i | \pi_j)}, \label{eq:crossfid}
\end{align}
where $e_i$ ($g_i$) denotes the assignment of \qb{j} to the excited (ground) state, $\pi_j$ ($0_j$) denotes the preparation with (without) a $\pi$-pulse on \qb{i} and the average $\exv{\cdot}$ is taken over assignment (preparation) of all qubits but $i$ ($j$).
We have extracted $F_{ij}$ from the experimental data presented in presented \cref{fig:probmatrix} and as shown in \cref{fig:crossfid} the off-diagonal elements are small.
Ideally, the outcome of \qb{i} should be uncorrelated with \qb{j} so we expect $P(e_i | 0_j) = P(g_i | \pi_j) = \num{0.5}$ and the off-diagonal elements should be \num{0}.
However, the correlations remain significantly below the individual readout infidelities obtained from the diagonal of $F_{ij}$ and thus we expect each weighted measurement band give information only about a single qubit. This is also in agreement with mode-matched filter having a very small spectral overlap. Single qubit errors therefore appears to dominate the imperfections of the qubit assignment.

The cross-fidelity $F_{ij}$ has a functional importance, as it shows if the state assignment of a qubit is affected by prepared the state of the others. However, as the readout threshold and mode-matched filter are optimized for the assignment of each qubit, a readout tone applied to qubit $i$ may carry additional information about qubit $j$ without causing imperfections visible in cross-fidelity $F_{ij}$. The complete physical influence on qubit $j$ from a readout pulse at qubit $i$ is characterized by the parasitic measurement-induced dephasing discussed in the \cref{sec:crosstalk}.

\begin{figure}
\begin{center}
\includegraphics[width=\linewidth]{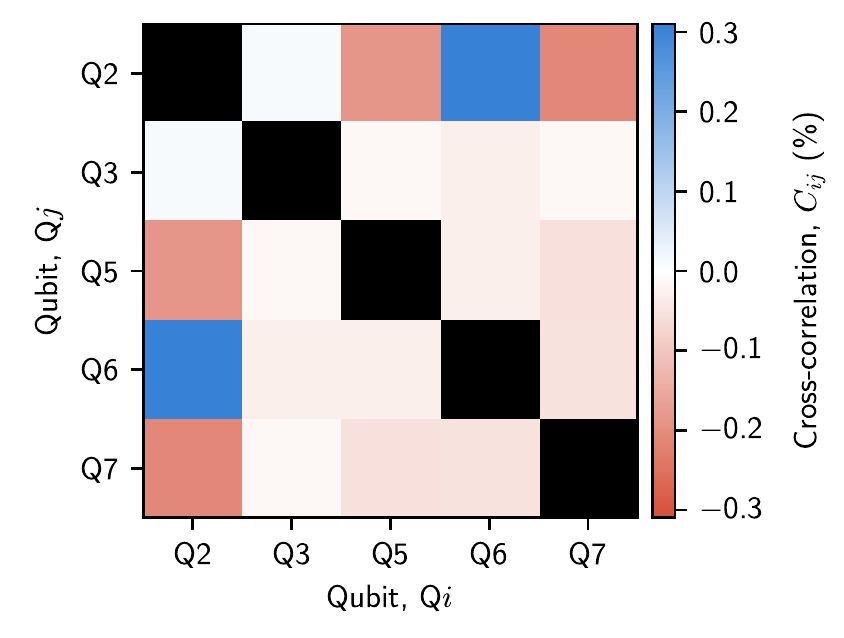}
\end{center}
\caption{Correlation coefficients between the outcomes of the multiplexed single-shot measurements as calculted by Eq.~\eqref{eq:covariance}} \label{fig:correlations}
\end{figure}

Another method to look for crosstalk in the data is to consider the cross-correlation matrix averaged over all prepared states
\begin{equation}
\begin{split}
C_{ij} &= \exv{\frac{\text{cov}\left(\sigma_{zi},\sigma_{zj}\right)}{ \sqrt{\text{var}\left(\sigma_{zi}\right) \text{var}\left(\sigma_{zj}\right)} }},
\end{split} \label{eq:covariance}
\end{equation}
shown in \cref{fig:correlations}. The positive (negative) cross-correlation indicates an over-representation of qubit pairs assigned to the same (opposite) state. As all states are prepared with an equal weight, we expect no cross-correlations while the diagonal elements $C_{ii}=1$ by definition. The off-diagonal elements of Fig~\ref{fig:correlations} are close to zero, thus we again see evidence that errors from correlations in the readout are significantly smaller than single-qubit errors. While correlation matrix indeed quantifies the crosstalk in the assignment, it has some drawbacks. For example in the trivial situation where the readout pulse would be turned off, the qubit state assignment would be highly correlated. Similarly, if the assignment of qubit $i$ to the exited state would always cause a bit-flip in the assignment of qubit $j$, the correlation-matrix element $C_{ij}$ would still be zero, as $C_{ij}$ is averaged over all qubit preparations. Thus the correlation matrix has limited applications for characterizing on-chip readout crosstalk.

\section{Measurement efficiency}
\label{app:efficiency}
The measurement-induced dephasing and the single-shot histograms give information about the measurement efficiency~\cite{Bultink2017}. The measurement-induced dephasing can be obtained from a Ramsey experiment with a measurement pulse interleaved between the two $\pi/2$-pulses also discussed in \cref{sec:crosstalk}. The average dephasing rates for each qubit are the diagonal elements, $\Gamma_{ii}$, of Fig.~\ref{fig:dephasing}, which are extracted by fitting a Gaussian to the Ramsey signal as a function of varying measurement pulse amplitude~\cite{Bultink2017}. In addition, from the single-shot histograms, such as the ones shown in \cref{fig:histograms}, we can extract the signal to noise ration (SNR), as the difference between the mean of the ground and excited state distributions divided by the standard deviation $\sigma$:
\begin{align}
\text{SNR} = \frac{\exv{s}_\pi - \exv{s}_0}{\sigma},
\end{align}
where $\exv{s}_{\pi (0)}$ is the average measured signal, $s$, when preparing the qubit with (without) a $\pi$-pulse prior to readout pulse and $\sigma$ is the standard deviation of the signal.

\begin{table}[tbh]
\centering
\begin{tabular*}{\linewidth}{l @{\extracolsep{\fill}} c c c c c}
\hline\hline
& Q2              & Q3              & Q5              & Q6              & Q7          \\ \hline
Efficiency, $\eta$ & 51.8\%      & 49.9\%      & 42.7\%      & 51.2\%      & 47.9\%   \\
\hline\hline
\end{tabular*}
\caption{The measurement efficiency for each qubit calculated by \cref{eq:eta} from a Ramsey type measurement discussed in the main text.} \label{tab:efficiencies}
\end{table}

As shown in Ref.~\cite{Bultink2017}, the quantum limited SNR can be expressed in terms of the integrated measurement induced dephasing, $\Gamma_{ii} \measlen$, such that
\begin{align}
\eta_i = \frac{\text{SNR}^2}{4 \Gamma_{ii} \measlen}. \label{eq:eta}
\end{align}
The observed measurement efficiencies are in the range of $42\%$ to $52\%$ as listed in \cref{tab:efficiencies}. The measurement efficiency is limited by the internal loss in the TWPA~\cite{Macklin2015} and in the MW components between the sample and TWPA. The differences of the measurement efficiency between each qubit may be explained by a frequency dependency in the gain of the TWPA and in the rest of the detection chain.

\bibliography{QudevRefDB}

\begin{thebibliography}{50}%
\makeatletter
\providecommand \@ifxundefined [1]{%
 \@ifx{#1\undefined}
}%
\providecommand \@ifnum [1]{%
 \ifnum #1\expandafter \@firstoftwo
 \else \expandafter \@secondoftwo
 \fi
}%
\providecommand \@ifx [1]{%
 \ifx #1\expandafter \@firstoftwo
 \else \expandafter \@secondoftwo
 \fi
}%
\providecommand \natexlab [1]{#1}%
\providecommand \enquote  [1]{``#1''}%
\providecommand \bibnamefont  [1]{#1}%
\providecommand \bibfnamefont [1]{#1}%
\providecommand \citenamefont [1]{#1}%
\providecommand \href@noop [0]{\@secondoftwo}%
\providecommand \href [0]{\begingroup \@sanitize@url \@href}%
\providecommand \@href[1]{\@@startlink{#1}\@@href}%
\providecommand \@@href[1]{\endgroup#1\@@endlink}%
\providecommand \@sanitize@url [0]{\catcode `\\12\catcode `\$12\catcode
  `\&12\catcode `\#12\catcode `\^12\catcode `\_12\catcode `\%12\relax}%
\providecommand \@@startlink[1]{}%
\providecommand \@@endlink[0]{}%
\providecommand \url  [0]{\begingroup\@sanitize@url \@url }%
\providecommand \@url [1]{\endgroup\@href {#1}{\urlprefix }}%
\providecommand \urlprefix  [0]{URL }%
\providecommand \Eprint [0]{\href }%
\providecommand \doibase [0]{http://dx.doi.org/}%
\providecommand \selectlanguage [0]{\@gobble}%
\providecommand \bibinfo  [0]{\@secondoftwo}%
\providecommand \bibfield  [0]{\@secondoftwo}%
\providecommand \translation [1]{[#1]}%
\providecommand \BibitemOpen [0]{}%
\providecommand \bibitemStop [0]{}%
\providecommand \bibitemNoStop [0]{.\EOS\space}%
\providecommand \EOS [0]{\spacefactor3000\relax}%
\providecommand \BibitemShut  [1]{\csname bibitem#1\endcsname}%
\let\auto@bib@innerbib\@empty
\bibitem [{\citenamefont {Nielsen}\ and\ \citenamefont
  {Chuang}(1997)}]{Nielsen1997}%
  \BibitemOpen
  \bibfield  {author} {\bibinfo {author} {\bibfnamefont {M.~A.}\ \bibnamefont
  {Nielsen}}\ and\ \bibinfo {author} {\bibfnamefont {I.~L.}\ \bibnamefont
  {Chuang}},\ }\href {\doibase 10.1103/PhysRevLett.79.321} {\bibfield
  {journal} {\bibinfo  {journal} {Phys. Rev. Lett.}\ }\textbf {\bibinfo
  {volume} {79}},\ \bibinfo {pages} {321} (\bibinfo {year} {1997})}\BibitemShut
  {NoStop}%
\bibitem [{\citenamefont {DiVincenzo}(2009)}]{DiVincenzo2009}%
  \BibitemOpen
  \bibfield  {author} {\bibinfo {author} {\bibfnamefont {D.~P.}\ \bibnamefont
  {DiVincenzo}},\ }\href {\doibase doi:10.1088/0031-8949/2009/T137/014020}
  {\bibfield  {journal} {\bibinfo  {journal} {Phys. Scr.}\ }\textbf {\bibinfo
  {volume} {2009}},\ \bibinfo {pages} {014020} (\bibinfo {year}
  {2009})}\BibitemShut {NoStop}%
\bibitem [{\citenamefont {Barends}\ \emph {et~al.}(2014)\citenamefont
  {Barends}, \citenamefont {Kelly}, \citenamefont {Megrant}, \citenamefont
  {Veitia}, \citenamefont {Sank}, \citenamefont {Jeffrey}, \citenamefont
  {White}, \citenamefont {Mutus}, \citenamefont {Fowler}, \citenamefont
  {Campbell}, \citenamefont {Chen}, \citenamefont {Chen}, \citenamefont
  {Chiaro}, \citenamefont {Dunsworth}, \citenamefont {Neill}, \citenamefont
  {{O\'Malley}}, \citenamefont {Roushan}, \citenamefont {Vainsencher},
  \citenamefont {Wenner}, \citenamefont {Korotkov}, \citenamefont {Cleland},\
  and\ \citenamefont {Martinis}}]{Barends2014}%
  \BibitemOpen
  \bibfield  {author} {\bibinfo {author} {\bibfnamefont {R.}~\bibnamefont
  {Barends}}, \bibinfo {author} {\bibfnamefont {J.}~\bibnamefont {Kelly}},
  \bibinfo {author} {\bibfnamefont {A.}~\bibnamefont {Megrant}}, \bibinfo
  {author} {\bibfnamefont {A.}~\bibnamefont {Veitia}}, \bibinfo {author}
  {\bibfnamefont {D.}~\bibnamefont {Sank}}, \bibinfo {author} {\bibfnamefont
  {E.}~\bibnamefont {Jeffrey}}, \bibinfo {author} {\bibfnamefont {T.~C.}\
  \bibnamefont {White}}, \bibinfo {author} {\bibfnamefont {J.}~\bibnamefont
  {Mutus}}, \bibinfo {author} {\bibfnamefont {A.~G.}\ \bibnamefont {Fowler}},
  \bibinfo {author} {\bibfnamefont {B.}~\bibnamefont {Campbell}}, \bibinfo
  {author} {\bibfnamefont {Y.}~\bibnamefont {Chen}}, \bibinfo {author}
  {\bibfnamefont {Z.}~\bibnamefont {Chen}}, \bibinfo {author} {\bibfnamefont
  {B.}~\bibnamefont {Chiaro}}, \bibinfo {author} {\bibfnamefont
  {A.}~\bibnamefont {Dunsworth}}, \bibinfo {author} {\bibfnamefont
  {C.}~\bibnamefont {Neill}}, \bibinfo {author} {\bibfnamefont
  {P.}~\bibnamefont {{O\'Malley}}}, \bibinfo {author} {\bibfnamefont
  {P.}~\bibnamefont {Roushan}}, \bibinfo {author} {\bibfnamefont
  {A.}~\bibnamefont {Vainsencher}}, \bibinfo {author} {\bibfnamefont
  {J.}~\bibnamefont {Wenner}}, \bibinfo {author} {\bibfnamefont {A.~N.}\
  \bibnamefont {Korotkov}}, \bibinfo {author} {\bibfnamefont {A.~N.}\
  \bibnamefont {Cleland}}, \ and\ \bibinfo {author} {\bibfnamefont {J.~M.}\
  \bibnamefont {Martinis}},\ }\href {\doibase 10.1038/nature13171} {\bibfield
  {journal} {\bibinfo  {journal} {Nature}\ }\textbf {\bibinfo {volume} {508}},\
  \bibinfo {pages} {500} (\bibinfo {year} {2014})}\BibitemShut {NoStop}%
\bibitem [{\citenamefont {Bennett}\ \emph {et~al.}(1993)\citenamefont
  {Bennett}, \citenamefont {Brassard}, \citenamefont {Cr\'epeau}, \citenamefont
  {Jozsa}, \citenamefont {Peres},\ and\ \citenamefont
  {Wootters}}]{Bennett1993}%
  \BibitemOpen
  \bibfield  {author} {\bibinfo {author} {\bibfnamefont {C.~H.}\ \bibnamefont
  {Bennett}}, \bibinfo {author} {\bibfnamefont {G.}~\bibnamefont {Brassard}},
  \bibinfo {author} {\bibfnamefont {C.}~\bibnamefont {Cr\'epeau}}, \bibinfo
  {author} {\bibfnamefont {R.}~\bibnamefont {Jozsa}}, \bibinfo {author}
  {\bibfnamefont {A.}~\bibnamefont {Peres}}, \ and\ \bibinfo {author}
  {\bibfnamefont {W.~K.}\ \bibnamefont {Wootters}},\ }\href {\doibase
  10.1103/PhysRevLett.70.1895} {\bibfield  {journal} {\bibinfo  {journal}
  {Phys. Rev. Lett.}\ }\textbf {\bibinfo {volume} {70}},\ \bibinfo {pages}
  {1895} (\bibinfo {year} {1993})}\BibitemShut {NoStop}%
\bibitem [{\citenamefont {Steffen}\ \emph {et~al.}(2013)\citenamefont
  {Steffen}, \citenamefont {Salathe}, \citenamefont {Oppliger}, \citenamefont
  {Kurpiers}, \citenamefont {Baur}, \citenamefont {Lang}, \citenamefont
  {Eichler}, \citenamefont {Puebla-Hellmann}, \citenamefont {Fedorov},\ and\
  \citenamefont {Wallraff}}]{Steffen2013}%
  \BibitemOpen
  \bibfield  {author} {\bibinfo {author} {\bibfnamefont {L.}~\bibnamefont
  {Steffen}}, \bibinfo {author} {\bibfnamefont {Y.}~\bibnamefont {Salathe}},
  \bibinfo {author} {\bibfnamefont {M.}~\bibnamefont {Oppliger}}, \bibinfo
  {author} {\bibfnamefont {P.}~\bibnamefont {Kurpiers}}, \bibinfo {author}
  {\bibfnamefont {M.}~\bibnamefont {Baur}}, \bibinfo {author} {\bibfnamefont
  {C.}~\bibnamefont {Lang}}, \bibinfo {author} {\bibfnamefont {C.}~\bibnamefont
  {Eichler}}, \bibinfo {author} {\bibfnamefont {G.}~\bibnamefont
  {Puebla-Hellmann}}, \bibinfo {author} {\bibfnamefont {A.}~\bibnamefont
  {Fedorov}}, \ and\ \bibinfo {author} {\bibfnamefont {A.}~\bibnamefont
  {Wallraff}},\ }\href {\doibase 10.1038/nature12422} {\bibfield  {journal}
  {\bibinfo  {journal} {Nature}\ }\textbf {\bibinfo {volume} {500}},\ \bibinfo
  {pages} {319} (\bibinfo {year} {2013})}\BibitemShut {NoStop}%
\bibitem [{\citenamefont {Johnson}\ \emph {et~al.}(2012)\citenamefont
  {Johnson}, \citenamefont {Macklin}, \citenamefont {Slichter}, \citenamefont
  {Vijay}, \citenamefont {Weingarten}, \citenamefont {Clarke},\ and\
  \citenamefont {Siddiqi}}]{Johnson2012}%
  \BibitemOpen
  \bibfield  {author} {\bibinfo {author} {\bibfnamefont {J.~E.}\ \bibnamefont
  {Johnson}}, \bibinfo {author} {\bibfnamefont {C.}~\bibnamefont {Macklin}},
  \bibinfo {author} {\bibfnamefont {D.~H.}\ \bibnamefont {Slichter}}, \bibinfo
  {author} {\bibfnamefont {R.}~\bibnamefont {Vijay}}, \bibinfo {author}
  {\bibfnamefont {E.~B.}\ \bibnamefont {Weingarten}}, \bibinfo {author}
  {\bibfnamefont {J.}~\bibnamefont {Clarke}}, \ and\ \bibinfo {author}
  {\bibfnamefont {I.}~\bibnamefont {Siddiqi}},\ }\href {\doibase
  10.1103/PhysRevLett.109.050506} {\bibfield  {journal} {\bibinfo  {journal}
  {Phys. Rev. Lett.}\ }\textbf {\bibinfo {volume} {109}},\ \bibinfo {pages}
  {050506} (\bibinfo {year} {2012})}\BibitemShut {NoStop}%
\bibitem [{\citenamefont {Rist\`e}\ \emph {et~al.}(2012)\citenamefont
  {Rist\`e}, \citenamefont {van Leeuwen}, \citenamefont {Ku}, \citenamefont
  {Lehnert},\ and\ \citenamefont {DiCarlo}}]{Riste2012}%
  \BibitemOpen
  \bibfield  {author} {\bibinfo {author} {\bibfnamefont {D.}~\bibnamefont
  {Rist\`e}}, \bibinfo {author} {\bibfnamefont {J.~G.}\ \bibnamefont {van
  Leeuwen}}, \bibinfo {author} {\bibfnamefont {H.-S.}\ \bibnamefont {Ku}},
  \bibinfo {author} {\bibfnamefont {K.~W.}\ \bibnamefont {Lehnert}}, \ and\
  \bibinfo {author} {\bibfnamefont {L.}~\bibnamefont {DiCarlo}},\ }\href
  {\doibase 10.1103/PhysRevLett.109.050507} {\bibfield  {journal} {\bibinfo
  {journal} {Phys. Rev. Lett.}\ }\textbf {\bibinfo {volume} {109}},\ \bibinfo
  {pages} {050507} (\bibinfo {year} {2012})}\BibitemShut {NoStop}%
\bibitem [{\citenamefont {Salath\'e}\ \emph {et~al.}(2017)\citenamefont
  {Salath\'e}, \citenamefont {Kurpiers}, \citenamefont {Karg}, \citenamefont
  {Lang}, \citenamefont {Andersen}, \citenamefont {Akin}, \citenamefont
  {Eichler},\ and\ \citenamefont {Wallraff}}]{Salathe2017}%
  \BibitemOpen
  \bibfield  {author} {\bibinfo {author} {\bibfnamefont {Y.}~\bibnamefont
  {Salath\'e}}, \bibinfo {author} {\bibfnamefont {P.}~\bibnamefont {Kurpiers}},
  \bibinfo {author} {\bibfnamefont {T.}~\bibnamefont {Karg}}, \bibinfo {author}
  {\bibfnamefont {C.}~\bibnamefont {Lang}}, \bibinfo {author} {\bibfnamefont
  {C.~K.}\ \bibnamefont {Andersen}}, \bibinfo {author} {\bibfnamefont
  {A.}~\bibnamefont {Akin}}, \bibinfo {author} {\bibfnamefont {C.}~\bibnamefont
  {Eichler}}, \ and\ \bibinfo {author} {\bibfnamefont {A.}~\bibnamefont
  {Wallraff}},\ }\href {https://arxiv.org/abs/1709.01030} {\bibfield  {journal}
  {\bibinfo  {journal} {arXiv:1709.01030}\ } (\bibinfo {year}
  {2017})}\BibitemShut {NoStop}%
\bibitem [{\citenamefont {Jerger}\ \emph {et~al.}(2012)\citenamefont {Jerger},
  \citenamefont {Poletto}, \citenamefont {Macha}, \citenamefont {Hübner},
  \citenamefont {Il’ichev},\ and\ \citenamefont {Ustinov}}]{Jerger2012}%
  \BibitemOpen
  \bibfield  {author} {\bibinfo {author} {\bibfnamefont {M.}~\bibnamefont
  {Jerger}}, \bibinfo {author} {\bibfnamefont {S.}~\bibnamefont {Poletto}},
  \bibinfo {author} {\bibfnamefont {P.}~\bibnamefont {Macha}}, \bibinfo
  {author} {\bibfnamefont {U.}~\bibnamefont {Hübner}}, \bibinfo {author}
  {\bibfnamefont {E.}~\bibnamefont {Il’ichev}}, \ and\ \bibinfo {author}
  {\bibfnamefont {A.~V.}\ \bibnamefont {Ustinov}},\ }\href {\doibase
  10.1063/1.4739454} {\bibfield  {journal} {\bibinfo  {journal} {Applied
  Physics Letters}\ }\textbf {\bibinfo {volume} {101}},\ \bibinfo {pages}
  {042604} (\bibinfo {year} {2012})},\ \Eprint
  {http://arxiv.org/abs/http://dx.doi.org/10.1063/1.4739454}
  {http://dx.doi.org/10.1063/1.4739454} \BibitemShut {NoStop}%
\bibitem [{\citenamefont {Schmitt}\ \emph {et~al.}(2014)\citenamefont
  {Schmitt}, \citenamefont {Zhou}, \citenamefont {Juliusson}, \citenamefont
  {Royer}, \citenamefont {Blais}, \citenamefont {Bertet}, \citenamefont
  {Vion},\ and\ \citenamefont {Esteve}}]{Schmitt2014a}%
  \BibitemOpen
  \bibfield  {author} {\bibinfo {author} {\bibfnamefont {V.}~\bibnamefont
  {Schmitt}}, \bibinfo {author} {\bibfnamefont {X.}~\bibnamefont {Zhou}},
  \bibinfo {author} {\bibfnamefont {K.}~\bibnamefont {Juliusson}}, \bibinfo
  {author} {\bibfnamefont {B.}~\bibnamefont {Royer}}, \bibinfo {author}
  {\bibfnamefont {A.}~\bibnamefont {Blais}}, \bibinfo {author} {\bibfnamefont
  {P.}~\bibnamefont {Bertet}}, \bibinfo {author} {\bibfnamefont
  {D.}~\bibnamefont {Vion}}, \ and\ \bibinfo {author} {\bibfnamefont
  {D.}~\bibnamefont {Esteve}},\ }\href {\doibase 10.1103/PhysRevA.90.062333}
  {\bibfield  {journal} {\bibinfo  {journal} {Phys. Rev. A}\ }\textbf {\bibinfo
  {volume} {90}},\ \bibinfo {pages} {062333} (\bibinfo {year}
  {2014})}\BibitemShut {NoStop}%
\bibitem [{\citenamefont {Jeffrey}\ \emph {et~al.}(2014)\citenamefont
  {Jeffrey}, \citenamefont {Sank}, \citenamefont {Mutus}, \citenamefont
  {White}, \citenamefont {Kelly}, \citenamefont {Barends}, \citenamefont
  {Chen}, \citenamefont {Chen}, \citenamefont {Chiaro}, \citenamefont
  {Dunsworth}, \citenamefont {Megrant}, \citenamefont {O'Malley}, \citenamefont
  {Neill}, \citenamefont {Roushan}, \citenamefont {Vainsencher}, \citenamefont
  {Wenner}, \citenamefont {Cleland},\ and\ \citenamefont
  {Martinis}}]{Jeffrey2014}%
  \BibitemOpen
  \bibfield  {author} {\bibinfo {author} {\bibfnamefont {E.}~\bibnamefont
  {Jeffrey}}, \bibinfo {author} {\bibfnamefont {D.}~\bibnamefont {Sank}},
  \bibinfo {author} {\bibfnamefont {J.~Y.}\ \bibnamefont {Mutus}}, \bibinfo
  {author} {\bibfnamefont {T.~C.}\ \bibnamefont {White}}, \bibinfo {author}
  {\bibfnamefont {J.}~\bibnamefont {Kelly}}, \bibinfo {author} {\bibfnamefont
  {R.}~\bibnamefont {Barends}}, \bibinfo {author} {\bibfnamefont
  {Y.}~\bibnamefont {Chen}}, \bibinfo {author} {\bibfnamefont {Z.}~\bibnamefont
  {Chen}}, \bibinfo {author} {\bibfnamefont {B.}~\bibnamefont {Chiaro}},
  \bibinfo {author} {\bibfnamefont {A.}~\bibnamefont {Dunsworth}}, \bibinfo
  {author} {\bibfnamefont {A.}~\bibnamefont {Megrant}}, \bibinfo {author}
  {\bibfnamefont {P.~J.~J.}\ \bibnamefont {O'Malley}}, \bibinfo {author}
  {\bibfnamefont {C.}~\bibnamefont {Neill}}, \bibinfo {author} {\bibfnamefont
  {P.}~\bibnamefont {Roushan}}, \bibinfo {author} {\bibfnamefont
  {A.}~\bibnamefont {Vainsencher}}, \bibinfo {author} {\bibfnamefont
  {J.}~\bibnamefont {Wenner}}, \bibinfo {author} {\bibfnamefont {A.~N.}\
  \bibnamefont {Cleland}}, \ and\ \bibinfo {author} {\bibfnamefont {J.~M.}\
  \bibnamefont {Martinis}},\ }\href {\doibase 10.1103/PhysRevLett.112.190504}
  {\bibfield  {journal} {\bibinfo  {journal} {Phys. Rev. Lett.}\ }\textbf
  {\bibinfo {volume} {112}},\ \bibinfo {pages} {190504} (\bibinfo {year}
  {2014})}\BibitemShut {NoStop}%
\bibitem [{\citenamefont {Blais}\ \emph {et~al.}(2004)\citenamefont {Blais},
  \citenamefont {Huang}, \citenamefont {Wallraff}, \citenamefont {Girvin},\
  and\ \citenamefont {Schoelkopf}}]{Blais2004}%
  \BibitemOpen
  \bibfield  {author} {\bibinfo {author} {\bibfnamefont {A.}~\bibnamefont
  {Blais}}, \bibinfo {author} {\bibfnamefont {R.-S.}\ \bibnamefont {Huang}},
  \bibinfo {author} {\bibfnamefont {A.}~\bibnamefont {Wallraff}}, \bibinfo
  {author} {\bibfnamefont {S.~M.}\ \bibnamefont {Girvin}}, \ and\ \bibinfo
  {author} {\bibfnamefont {R.~J.}\ \bibnamefont {Schoelkopf}},\ }\href
  {\doibase 10.1103/PhysRevA.69.062320} {\bibfield  {journal} {\bibinfo
  {journal} {Phys. Rev. A}\ }\textbf {\bibinfo {volume} {69}},\ \bibinfo
  {pages} {062320} (\bibinfo {year} {2004})}\BibitemShut {NoStop}%
\bibitem [{\citenamefont {Wallraff}\ \emph {et~al.}(2005)\citenamefont
  {Wallraff}, \citenamefont {Schuster}, \citenamefont {Blais}, \citenamefont
  {Frunzio}, \citenamefont {Majer}, \citenamefont {Devoret}, \citenamefont
  {Girvin},\ and\ \citenamefont {Schoelkopf}}]{Wallraff2005}%
  \BibitemOpen
  \bibfield  {author} {\bibinfo {author} {\bibfnamefont {A.}~\bibnamefont
  {Wallraff}}, \bibinfo {author} {\bibfnamefont {D.~I.}\ \bibnamefont
  {Schuster}}, \bibinfo {author} {\bibfnamefont {A.}~\bibnamefont {Blais}},
  \bibinfo {author} {\bibfnamefont {L.}~\bibnamefont {Frunzio}}, \bibinfo
  {author} {\bibfnamefont {J.}~\bibnamefont {Majer}}, \bibinfo {author}
  {\bibfnamefont {M.~H.}\ \bibnamefont {Devoret}}, \bibinfo {author}
  {\bibfnamefont {S.~M.}\ \bibnamefont {Girvin}}, \ and\ \bibinfo {author}
  {\bibfnamefont {R.~J.}\ \bibnamefont {Schoelkopf}},\ }\href {\doibase
  10.1103/PhysRevLett.95.060501} {\bibfield  {journal} {\bibinfo  {journal}
  {Phys. Rev. Lett.}\ }\textbf {\bibinfo {volume} {95}},\ \bibinfo {pages}
  {060501} (\bibinfo {year} {2005})}\BibitemShut {NoStop}%
\bibitem [{\citenamefont {Caves}(1982)}]{Caves1982}%
  \BibitemOpen
  \bibfield  {author} {\bibinfo {author} {\bibfnamefont {C.~M.}\ \bibnamefont
  {Caves}},\ }\href {\doibase 10.1103/PhysRevD.26.1817} {\bibfield  {journal}
  {\bibinfo  {journal} {Phys. Rev. D}\ }\textbf {\bibinfo {volume} {26}},\
  \bibinfo {pages} {1817} (\bibinfo {year} {1982})}\BibitemShut {NoStop}%
\bibitem [{\citenamefont {Yurke}\ \emph {et~al.}(1996)\citenamefont {Yurke},
  \citenamefont {Roukes}, \citenamefont {Movshovich},\ and\ \citenamefont
  {Pargellis}}]{Yurke1996}%
  \BibitemOpen
  \bibfield  {author} {\bibinfo {author} {\bibfnamefont {B.}~\bibnamefont
  {Yurke}}, \bibinfo {author} {\bibfnamefont {M.~L.}\ \bibnamefont {Roukes}},
  \bibinfo {author} {\bibfnamefont {R.}~\bibnamefont {Movshovich}}, \ and\
  \bibinfo {author} {\bibfnamefont {A.~N.}\ \bibnamefont {Pargellis}},\ }\href
  {http://dx.doi.org/10.1063/1.116845} {\bibfield  {journal} {\bibinfo
  {journal} {Appl. Phys. Lett.}\ }\textbf {\bibinfo {volume} {69}},\ \bibinfo
  {pages} {3078} (\bibinfo {year} {1996})}\BibitemShut {NoStop}%
\bibitem [{\citenamefont {Castellanos-Beltran}\ \emph
  {et~al.}(2008)\citenamefont {Castellanos-Beltran}, \citenamefont {Irwin},
  \citenamefont {Hilton}, \citenamefont {Vale},\ and\ \citenamefont
  {Lehnert}}]{Castellanos2008}%
  \BibitemOpen
  \bibfield  {author} {\bibinfo {author} {\bibfnamefont {M.~A.}\ \bibnamefont
  {Castellanos-Beltran}}, \bibinfo {author} {\bibfnamefont {K.~D.}\
  \bibnamefont {Irwin}}, \bibinfo {author} {\bibfnamefont {G.~C.}\ \bibnamefont
  {Hilton}}, \bibinfo {author} {\bibfnamefont {L.~R.}\ \bibnamefont {Vale}}, \
  and\ \bibinfo {author} {\bibfnamefont {K.~W.}\ \bibnamefont {Lehnert}},\
  }\href {\doibase 10.1038/nphys1090} {\bibfield  {journal} {\bibinfo
  {journal} {Nat. Phys.}\ }\textbf {\bibinfo {volume} {4}},\ \bibinfo {pages}
  {929} (\bibinfo {year} {2008})}\BibitemShut {NoStop}%
\bibitem [{\citenamefont {Eichler}\ \emph {et~al.}(2014)\citenamefont
  {Eichler}, \citenamefont {Salathe}, \citenamefont {Mlynek}, \citenamefont
  {Schmidt},\ and\ \citenamefont {Wallraff}}]{Eichler2014a}%
  \BibitemOpen
  \bibfield  {author} {\bibinfo {author} {\bibfnamefont {C.}~\bibnamefont
  {Eichler}}, \bibinfo {author} {\bibfnamefont {Y.}~\bibnamefont {Salathe}},
  \bibinfo {author} {\bibfnamefont {J.}~\bibnamefont {Mlynek}}, \bibinfo
  {author} {\bibfnamefont {S.}~\bibnamefont {Schmidt}}, \ and\ \bibinfo
  {author} {\bibfnamefont {A.}~\bibnamefont {Wallraff}},\ }\href {\doibase
  10.1103/PhysRevLett.113.110502} {\bibfield  {journal} {\bibinfo  {journal}
  {Phys. Rev. Lett.}\ }\textbf {\bibinfo {volume} {113}},\ \bibinfo {pages}
  {110502} (\bibinfo {year} {2014})}\BibitemShut {NoStop}%
\bibitem [{\citenamefont {Mallet}\ \emph {et~al.}(2009)\citenamefont {Mallet},
  \citenamefont {Ong}, \citenamefont {Palacios-Laloy}, \citenamefont {Nguyen},
  \citenamefont {Bertet}, \citenamefont {Vion},\ and\ \citenamefont
  {Esteve}}]{Mallet2009}%
  \BibitemOpen
  \bibfield  {author} {\bibinfo {author} {\bibfnamefont {F.}~\bibnamefont
  {Mallet}}, \bibinfo {author} {\bibfnamefont {F.~R.}\ \bibnamefont {Ong}},
  \bibinfo {author} {\bibfnamefont {A.}~\bibnamefont {Palacios-Laloy}},
  \bibinfo {author} {\bibfnamefont {F.}~\bibnamefont {Nguyen}}, \bibinfo
  {author} {\bibfnamefont {P.}~\bibnamefont {Bertet}}, \bibinfo {author}
  {\bibfnamefont {D.}~\bibnamefont {Vion}}, \ and\ \bibinfo {author}
  {\bibfnamefont {D.}~\bibnamefont {Esteve}},\ }\href {\doibase
  10.1038/nphys1400} {\bibfield  {journal} {\bibinfo  {journal} {Nat. Phys.}\
  }\textbf {\bibinfo {volume} {5}},\ \bibinfo {pages} {791} (\bibinfo {year}
  {2009})}\BibitemShut {NoStop}%
\bibitem [{\citenamefont {Vijay}\ \emph {et~al.}(2011)\citenamefont {Vijay},
  \citenamefont {Slichter},\ and\ \citenamefont {Siddiqi}}]{Vijay2011}%
  \BibitemOpen
  \bibfield  {author} {\bibinfo {author} {\bibfnamefont {R.}~\bibnamefont
  {Vijay}}, \bibinfo {author} {\bibfnamefont {D.~H.}\ \bibnamefont {Slichter}},
  \ and\ \bibinfo {author} {\bibfnamefont {I.}~\bibnamefont {Siddiqi}},\ }\href
  {\doibase 10.1103/PhysRevLett.106.110502} {\bibfield  {journal} {\bibinfo
  {journal} {Phys. Rev. Lett.}\ }\textbf {\bibinfo {volume} {106}},\ \bibinfo
  {pages} {110502} (\bibinfo {year} {2011})}\BibitemShut {NoStop}%
\bibitem [{\citenamefont {Reed}\ \emph {et~al.}(2010)\citenamefont {Reed},
  \citenamefont {Johnson}, \citenamefont {Houck}, \citenamefont {DiCarlo},
  \citenamefont {Chow}, \citenamefont {Schuster}, \citenamefont {Frunzio},\
  and\ \citenamefont {Schoelkopf}}]{Reed2010}%
  \BibitemOpen
  \bibfield  {author} {\bibinfo {author} {\bibfnamefont {M.~D.}\ \bibnamefont
  {Reed}}, \bibinfo {author} {\bibfnamefont {B.~R.}\ \bibnamefont {Johnson}},
  \bibinfo {author} {\bibfnamefont {A.~A.}\ \bibnamefont {Houck}}, \bibinfo
  {author} {\bibfnamefont {L.}~\bibnamefont {DiCarlo}}, \bibinfo {author}
  {\bibfnamefont {J.~M.}\ \bibnamefont {Chow}}, \bibinfo {author}
  {\bibfnamefont {D.~I.}\ \bibnamefont {Schuster}}, \bibinfo {author}
  {\bibfnamefont {L.}~\bibnamefont {Frunzio}}, \ and\ \bibinfo {author}
  {\bibfnamefont {R.~J.}\ \bibnamefont {Schoelkopf}},\ }\href {\doibase
  10.1063/1.3435463} {\bibfield  {journal} {\bibinfo  {journal} {Appl. Phys.
  Lett.}\ }\textbf {\bibinfo {volume} {96}},\ \bibinfo {eid} {203110} (\bibinfo
  {year} {2010})}\BibitemShut {NoStop}%
\bibitem [{\citenamefont {Bronn}\ \emph {et~al.}(2015)\citenamefont {Bronn},
  \citenamefont {Liu}, \citenamefont {Hertzberg}, \citenamefont {C\'{o}rcoles},
  \citenamefont {Houck}, \citenamefont {Gambetta},\ and\ \citenamefont
  {Chow}}]{Bronn2015b}%
  \BibitemOpen
  \bibfield  {author} {\bibinfo {author} {\bibfnamefont {N.~T.}\ \bibnamefont
  {Bronn}}, \bibinfo {author} {\bibfnamefont {Y.}~\bibnamefont {Liu}}, \bibinfo
  {author} {\bibfnamefont {J.~B.}\ \bibnamefont {Hertzberg}}, \bibinfo {author}
  {\bibfnamefont {A.~D.}\ \bibnamefont {C\'{o}rcoles}}, \bibinfo {author}
  {\bibfnamefont {A.~A.}\ \bibnamefont {Houck}}, \bibinfo {author}
  {\bibfnamefont {J.~M.}\ \bibnamefont {Gambetta}}, \ and\ \bibinfo {author}
  {\bibfnamefont {J.~M.}\ \bibnamefont {Chow}},\ }\href {\doibase
  http://dx.doi.org/10.1063/1.4934867} {\bibfield  {journal} {\bibinfo
  {journal} {Applied Physics Letters}\ }\textbf {\bibinfo {volume} {107}},\
  \bibinfo {eid} {172601} (\bibinfo {year} {2015})}\BibitemShut {NoStop}%
\bibitem [{\citenamefont {Walter}\ \emph {et~al.}(2017)\citenamefont {Walter},
  \citenamefont {Kurpiers}, \citenamefont {Gasparinetti}, \citenamefont
  {Magnard}, \citenamefont {Potocnik}, \citenamefont {Salath\'e}, \citenamefont
  {Pechal}, \citenamefont {Mondal}, \citenamefont {Oppliger}, \citenamefont
  {Eichler},\ and\ \citenamefont {Wallraff}}]{Walter2017}%
  \BibitemOpen
  \bibfield  {author} {\bibinfo {author} {\bibfnamefont {T.}~\bibnamefont
  {Walter}}, \bibinfo {author} {\bibfnamefont {P.}~\bibnamefont {Kurpiers}},
  \bibinfo {author} {\bibfnamefont {S.}~\bibnamefont {Gasparinetti}}, \bibinfo
  {author} {\bibfnamefont {P.}~\bibnamefont {Magnard}}, \bibinfo {author}
  {\bibfnamefont {A.}~\bibnamefont {Potocnik}}, \bibinfo {author}
  {\bibfnamefont {Y.}~\bibnamefont {Salath\'e}}, \bibinfo {author}
  {\bibfnamefont {M.}~\bibnamefont {Pechal}}, \bibinfo {author} {\bibfnamefont
  {M.}~\bibnamefont {Mondal}}, \bibinfo {author} {\bibfnamefont
  {M.}~\bibnamefont {Oppliger}}, \bibinfo {author} {\bibfnamefont
  {C.}~\bibnamefont {Eichler}}, \ and\ \bibinfo {author} {\bibfnamefont
  {A.}~\bibnamefont {Wallraff}},\ }\href {\doibase
  10.1103/PhysRevApplied.7.054020} {\bibfield  {journal} {\bibinfo  {journal}
  {Phys. Rev. Applied}\ }\textbf {\bibinfo {volume} {7}},\ \bibinfo {pages}
  {054020} (\bibinfo {year} {2017})}\BibitemShut {NoStop}%
\bibitem [{\citenamefont {Filipp}\ \emph {et~al.}(2009)\citenamefont {Filipp},
  \citenamefont {Maurer}, \citenamefont {Leek}, \citenamefont {Baur},
  \citenamefont {Bianchetti}, \citenamefont {Fink}, \citenamefont {G\"{o}ppl},
  \citenamefont {Steffen}, \citenamefont {Gambetta}, \citenamefont {Blais},\
  and\ \citenamefont {Wallraff}}]{Filipp2009b}%
  \BibitemOpen
  \bibfield  {author} {\bibinfo {author} {\bibfnamefont {S.}~\bibnamefont
  {Filipp}}, \bibinfo {author} {\bibfnamefont {P.}~\bibnamefont {Maurer}},
  \bibinfo {author} {\bibfnamefont {P.~J.}\ \bibnamefont {Leek}}, \bibinfo
  {author} {\bibfnamefont {M.}~\bibnamefont {Baur}}, \bibinfo {author}
  {\bibfnamefont {R.}~\bibnamefont {Bianchetti}}, \bibinfo {author}
  {\bibfnamefont {J.~M.}\ \bibnamefont {Fink}}, \bibinfo {author}
  {\bibfnamefont {M.}~\bibnamefont {G\"{o}ppl}}, \bibinfo {author}
  {\bibfnamefont {L.}~\bibnamefont {Steffen}}, \bibinfo {author} {\bibfnamefont
  {J.~M.}\ \bibnamefont {Gambetta}}, \bibinfo {author} {\bibfnamefont
  {A.}~\bibnamefont {Blais}}, \ and\ \bibinfo {author} {\bibfnamefont
  {A.}~\bibnamefont {Wallraff}},\ }\href {\doibase
  10.1103/PhysRevLett.102.200402} {\bibfield  {journal} {\bibinfo  {journal}
  {Phys. Rev. Lett.}\ }\textbf {\bibinfo {volume} {102}},\ \bibinfo {eid}
  {200402} (\bibinfo {year} {2009})}\BibitemShut {NoStop}%
\bibitem [{\citenamefont {DiCarlo}\ \emph {et~al.}(2010)\citenamefont
  {DiCarlo}, \citenamefont {Reed}, \citenamefont {Sun}, \citenamefont
  {Johnson}, \citenamefont {Chow}, \citenamefont {Gambetta}, \citenamefont
  {Frunzio}, \citenamefont {Girvin}, \citenamefont {Devoret},\ and\
  \citenamefont {Schoelkopf}}]{DiCarlo2010}%
  \BibitemOpen
  \bibfield  {author} {\bibinfo {author} {\bibfnamefont {L.}~\bibnamefont
  {DiCarlo}}, \bibinfo {author} {\bibfnamefont {M.~D.}\ \bibnamefont {Reed}},
  \bibinfo {author} {\bibfnamefont {L.}~\bibnamefont {Sun}}, \bibinfo {author}
  {\bibfnamefont {B.~R.}\ \bibnamefont {Johnson}}, \bibinfo {author}
  {\bibfnamefont {J.~M.}\ \bibnamefont {Chow}}, \bibinfo {author}
  {\bibfnamefont {J.~M.}\ \bibnamefont {Gambetta}}, \bibinfo {author}
  {\bibfnamefont {L.}~\bibnamefont {Frunzio}}, \bibinfo {author} {\bibfnamefont
  {S.~M.}\ \bibnamefont {Girvin}}, \bibinfo {author} {\bibfnamefont {M.~H.}\
  \bibnamefont {Devoret}}, \ and\ \bibinfo {author} {\bibfnamefont {R.~J.}\
  \bibnamefont {Schoelkopf}},\ }\href {\doibase doi:10.1038/nature09416}
  {\bibfield  {journal} {\bibinfo  {journal} {Nature}\ }\textbf {\bibinfo
  {volume} {467}},\ \bibinfo {pages} {574} (\bibinfo {year}
  {2010})}\BibitemShut {NoStop}%
\bibitem [{\citenamefont {Mutus}\ \emph {et~al.}(2014)\citenamefont {Mutus},
  \citenamefont {White}, \citenamefont {Barends}, \citenamefont {Chen},
  \citenamefont {Chen}, \citenamefont {Chiaro}, \citenamefont {Dunsworth},
  \citenamefont {Jeffrey}, \citenamefont {Kelly}, \citenamefont {Megrant},
  \citenamefont {Neill}, \citenamefont {O'Malley}, \citenamefont {Roushan},
  \citenamefont {Sank}, \citenamefont {Vainsencher}, \citenamefont {Wenner},
  \citenamefont {Sundqvist}, \citenamefont {Cleland},\ and\ \citenamefont
  {Martinis}}]{Mutus2014a}%
  \BibitemOpen
  \bibfield  {author} {\bibinfo {author} {\bibfnamefont {J.~Y.}\ \bibnamefont
  {Mutus}}, \bibinfo {author} {\bibfnamefont {T.~C.}\ \bibnamefont {White}},
  \bibinfo {author} {\bibfnamefont {R.}~\bibnamefont {Barends}}, \bibinfo
  {author} {\bibfnamefont {Y.}~\bibnamefont {Chen}}, \bibinfo {author}
  {\bibfnamefont {Z.}~\bibnamefont {Chen}}, \bibinfo {author} {\bibfnamefont
  {B.}~\bibnamefont {Chiaro}}, \bibinfo {author} {\bibfnamefont
  {A.}~\bibnamefont {Dunsworth}}, \bibinfo {author} {\bibfnamefont
  {E.}~\bibnamefont {Jeffrey}}, \bibinfo {author} {\bibfnamefont
  {J.}~\bibnamefont {Kelly}}, \bibinfo {author} {\bibfnamefont
  {A.}~\bibnamefont {Megrant}}, \bibinfo {author} {\bibfnamefont
  {C.}~\bibnamefont {Neill}}, \bibinfo {author} {\bibfnamefont {P.~J.~J.}\
  \bibnamefont {O'Malley}}, \bibinfo {author} {\bibfnamefont {P.}~\bibnamefont
  {Roushan}}, \bibinfo {author} {\bibfnamefont {D.}~\bibnamefont {Sank}},
  \bibinfo {author} {\bibfnamefont {A.}~\bibnamefont {Vainsencher}}, \bibinfo
  {author} {\bibfnamefont {J.}~\bibnamefont {Wenner}}, \bibinfo {author}
  {\bibfnamefont {K.~M.}\ \bibnamefont {Sundqvist}}, \bibinfo {author}
  {\bibfnamefont {A.~N.}\ \bibnamefont {Cleland}}, \ and\ \bibinfo {author}
  {\bibfnamefont {J.~M.}\ \bibnamefont {Martinis}},\ }\href {\doibase
  http://dx.doi.org/10.1063/1.4886408} {\bibfield  {journal} {\bibinfo
  {journal} {Applied Physics Letters}\ }\textbf {\bibinfo {volume} {104}},\
  \bibinfo {eid} {263513} (\bibinfo {year} {2014})}\BibitemShut {NoStop}%
\bibitem [{\citenamefont {Macklin}\ \emph {et~al.}(2015)\citenamefont
  {Macklin}, \citenamefont {O'Brien}, \citenamefont {Hover}, \citenamefont
  {Schwartz}, \citenamefont {Bolkhovsky}, \citenamefont {Zhang}, \citenamefont
  {Oliver},\ and\ \citenamefont {Siddiqi}}]{Macklin2015}%
  \BibitemOpen
  \bibfield  {author} {\bibinfo {author} {\bibfnamefont {C.}~\bibnamefont
  {Macklin}}, \bibinfo {author} {\bibfnamefont {K.}~\bibnamefont {O'Brien}},
  \bibinfo {author} {\bibfnamefont {D.}~\bibnamefont {Hover}}, \bibinfo
  {author} {\bibfnamefont {M.~E.}\ \bibnamefont {Schwartz}}, \bibinfo {author}
  {\bibfnamefont {V.}~\bibnamefont {Bolkhovsky}}, \bibinfo {author}
  {\bibfnamefont {X.}~\bibnamefont {Zhang}}, \bibinfo {author} {\bibfnamefont
  {W.~D.}\ \bibnamefont {Oliver}}, \ and\ \bibinfo {author} {\bibfnamefont
  {I.}~\bibnamefont {Siddiqi}},\ }\href {\doibase 10.1126/science.aaa8525}
  {\bibfield  {journal} {\bibinfo  {journal} {Science}\ }\textbf {\bibinfo
  {volume} {350}},\ \bibinfo {pages} {307} (\bibinfo {year}
  {2015})}\BibitemShut {NoStop}%
\bibitem [{\citenamefont {Roy}\ \emph {et~al.}(2015)\citenamefont {Roy},
  \citenamefont {Kundu}, \citenamefont {Chand}, \citenamefont {Vadiraj},
  \citenamefont {Ranadive}, \citenamefont {Nehra}, \citenamefont {Patankar},
  \citenamefont {Aumentado}, \citenamefont {Clerk},\ and\ \citenamefont
  {Vijay}}]{Roy2015c}%
  \BibitemOpen
  \bibfield  {author} {\bibinfo {author} {\bibfnamefont {T.}~\bibnamefont
  {Roy}}, \bibinfo {author} {\bibfnamefont {S.}~\bibnamefont {Kundu}}, \bibinfo
  {author} {\bibfnamefont {M.}~\bibnamefont {Chand}}, \bibinfo {author}
  {\bibfnamefont {A.~M.}\ \bibnamefont {Vadiraj}}, \bibinfo {author}
  {\bibfnamefont {A.}~\bibnamefont {Ranadive}}, \bibinfo {author}
  {\bibfnamefont {N.}~\bibnamefont {Nehra}}, \bibinfo {author} {\bibfnamefont
  {M.~P.}\ \bibnamefont {Patankar}}, \bibinfo {author} {\bibfnamefont
  {J.}~\bibnamefont {Aumentado}}, \bibinfo {author} {\bibfnamefont {A.~A.}\
  \bibnamefont {Clerk}}, \ and\ \bibinfo {author} {\bibfnamefont
  {R.}~\bibnamefont {Vijay}},\ }\href {\doibase 10.1063/1.4939148} {\bibfield
  {journal} {\bibinfo  {journal} {Appl. Phys. Lett.}\ }\textbf {\bibinfo
  {volume} {107}},\ \bibinfo {pages} {262601} (\bibinfo {year}
  {2015})}\BibitemShut {NoStop}%
\bibitem [{\citenamefont {Neill}\ \emph {et~al.}(2017)\citenamefont {Neill},
  \citenamefont {Roushan}, \citenamefont {Kechedzhi}, \citenamefont {Boixo},
  \citenamefont {Isakov}, \citenamefont {Smelyanskiy}, \citenamefont {Barends},
  \citenamefont {Burkett}, \citenamefont {Chen}, \citenamefont {Chen},
  \citenamefont {Chiaro}, \citenamefont {Dunsworth}, \citenamefont {Fowler},
  \citenamefont {Foxen}, \citenamefont {Graff}, \citenamefont {Jeffrey},
  \citenamefont {Kelly}, \citenamefont {Lucero}, \citenamefont {Megrant},
  \citenamefont {Mutus}, \citenamefont {Neeley}, \citenamefont {Quintana},
  \citenamefont {Sank}, \citenamefont {Vainsencher}, \citenamefont {Wenner},
  \citenamefont {White}, \citenamefont {Neven},\ and\ \citenamefont
  {Martinis}}]{Neill2017}%
  \BibitemOpen
  \bibfield  {author} {\bibinfo {author} {\bibfnamefont {C.}~\bibnamefont
  {Neill}}, \bibinfo {author} {\bibfnamefont {P.}~\bibnamefont {Roushan}},
  \bibinfo {author} {\bibfnamefont {K.}~\bibnamefont {Kechedzhi}}, \bibinfo
  {author} {\bibfnamefont {S.}~\bibnamefont {Boixo}}, \bibinfo {author}
  {\bibfnamefont {S.~V.}\ \bibnamefont {Isakov}}, \bibinfo {author}
  {\bibfnamefont {V.}~\bibnamefont {Smelyanskiy}}, \bibinfo {author}
  {\bibfnamefont {R.}~\bibnamefont {Barends}}, \bibinfo {author} {\bibfnamefont
  {B.}~\bibnamefont {Burkett}}, \bibinfo {author} {\bibfnamefont
  {Y.}~\bibnamefont {Chen}}, \bibinfo {author} {\bibfnamefont {Z.}~\bibnamefont
  {Chen}}, \bibinfo {author} {\bibfnamefont {B.}~\bibnamefont {Chiaro}},
  \bibinfo {author} {\bibfnamefont {A.}~\bibnamefont {Dunsworth}}, \bibinfo
  {author} {\bibfnamefont {A.}~\bibnamefont {Fowler}}, \bibinfo {author}
  {\bibfnamefont {B.}~\bibnamefont {Foxen}}, \bibinfo {author} {\bibfnamefont
  {R.}~\bibnamefont {Graff}}, \bibinfo {author} {\bibfnamefont
  {E.}~\bibnamefont {Jeffrey}}, \bibinfo {author} {\bibfnamefont
  {J.}~\bibnamefont {Kelly}}, \bibinfo {author} {\bibfnamefont
  {E.}~\bibnamefont {Lucero}}, \bibinfo {author} {\bibfnamefont
  {A.}~\bibnamefont {Megrant}}, \bibinfo {author} {\bibfnamefont
  {J.}~\bibnamefont {Mutus}}, \bibinfo {author} {\bibfnamefont
  {M.}~\bibnamefont {Neeley}}, \bibinfo {author} {\bibfnamefont
  {C.}~\bibnamefont {Quintana}}, \bibinfo {author} {\bibfnamefont
  {D.}~\bibnamefont {Sank}}, \bibinfo {author} {\bibfnamefont {A.}~\bibnamefont
  {Vainsencher}}, \bibinfo {author} {\bibfnamefont {J.}~\bibnamefont {Wenner}},
  \bibinfo {author} {\bibfnamefont {T.~C.}\ \bibnamefont {White}}, \bibinfo
  {author} {\bibfnamefont {H.}~\bibnamefont {Neven}}, \ and\ \bibinfo {author}
  {\bibfnamefont {J.~M.}\ \bibnamefont {Martinis}},\ }\href
  {https://arxiv.org/abs/1709.06678} {\bibfield  {journal} {\bibinfo  {journal}
  {arXiv:1709.06678}\ } (\bibinfo {year} {2017})}\BibitemShut {NoStop}%
\bibitem [{\citenamefont {Bronn}\ \emph {et~al.}(2017)\citenamefont {Bronn},
  \citenamefont {Abdo}, \citenamefont {Inoue}, \citenamefont {Lekuch},
  \citenamefont {Córcoles}, \citenamefont {Hertzberg}, \citenamefont {Takita},
  \citenamefont {Bishop}, \citenamefont {Gambetta},\ and\ \citenamefont
  {Chow}}]{Bronn2017a}%
  \BibitemOpen
  \bibfield  {author} {\bibinfo {author} {\bibfnamefont {N.~T.}\ \bibnamefont
  {Bronn}}, \bibinfo {author} {\bibfnamefont {B.}~\bibnamefont {Abdo}},
  \bibinfo {author} {\bibfnamefont {K.}~\bibnamefont {Inoue}}, \bibinfo
  {author} {\bibfnamefont {S.}~\bibnamefont {Lekuch}}, \bibinfo {author}
  {\bibfnamefont {A.~D.}\ \bibnamefont {Córcoles}}, \bibinfo {author}
  {\bibfnamefont {J.~B.}\ \bibnamefont {Hertzberg}}, \bibinfo {author}
  {\bibfnamefont {M.}~\bibnamefont {Takita}}, \bibinfo {author} {\bibfnamefont
  {L.~S.}\ \bibnamefont {Bishop}}, \bibinfo {author} {\bibfnamefont {J.~M.}\
  \bibnamefont {Gambetta}}, \ and\ \bibinfo {author} {\bibfnamefont {J.~M.}\
  \bibnamefont {Chow}},\ }\href
  {http://stacks.iop.org/1742-6596/834/i=1/a=012003} {\bibfield  {journal}
  {\bibinfo  {journal} {Journal of Physics: Conference Series}\ }\textbf
  {\bibinfo {volume} {834}},\ \bibinfo {pages} {012003} (\bibinfo {year}
  {2017})}\BibitemShut {NoStop}%
\bibitem [{\citenamefont {Corcoles}\ \emph {et~al.}(2015)\citenamefont
  {Corcoles}, \citenamefont {Magesan}, \citenamefont {Srinivasan},
  \citenamefont {Cross}, \citenamefont {Steffen}, \citenamefont {Gambetta},\
  and\ \citenamefont {Chow}}]{Corcoles2015}%
  \BibitemOpen
  \bibfield  {author} {\bibinfo {author} {\bibfnamefont {A.}~\bibnamefont
  {Corcoles}}, \bibinfo {author} {\bibfnamefont {E.}~\bibnamefont {Magesan}},
  \bibinfo {author} {\bibfnamefont {S.~J.}\ \bibnamefont {Srinivasan}},
  \bibinfo {author} {\bibfnamefont {A.~W.}\ \bibnamefont {Cross}}, \bibinfo
  {author} {\bibfnamefont {M.}~\bibnamefont {Steffen}}, \bibinfo {author}
  {\bibfnamefont {J.~M.}\ \bibnamefont {Gambetta}}, \ and\ \bibinfo {author}
  {\bibfnamefont {J.~M.}\ \bibnamefont {Chow}},\ }\href
  {http://dx.doi.org/10.1038/ncomms7979} {\bibfield  {journal} {\bibinfo
  {journal} {Nat Commun}\ }\textbf {\bibinfo {volume} {6}},\  (\bibinfo {year}
  {2015})}\BibitemShut {NoStop}%
\bibitem [{\citenamefont {Reagor}\ \emph {et~al.}(2017)\citenamefont {Reagor},
  \citenamefont {Osborn}, \citenamefont {Tezak}, \citenamefont {Staley},
  \citenamefont {Prawiroatmodjo}, \citenamefont {Scheer}, \citenamefont
  {Alidoust}, \citenamefont {Sete}, \citenamefont {Didier}, \citenamefont
  {da~Silva}, \citenamefont {Acala}, \citenamefont {Angeles}, \citenamefont
  {Bestwick}, \citenamefont {Block}, \citenamefont {Bloom}, \citenamefont
  {Bradley}, \citenamefont {Bui}, \citenamefont {Caldwell}, \citenamefont
  {Capelluto}, \citenamefont {Chilcott}, \citenamefont {Cordova}, \citenamefont
  {Crossman}, \citenamefont {Curtis}, \citenamefont {Deshpande}, \citenamefont
  {Bouayadi}, \citenamefont {Girshovich}, \citenamefont {Hong}, \citenamefont
  {Hudson}, \citenamefont {Karalekas}, \citenamefont {Kuang}, \citenamefont
  {Lenihan}, \citenamefont {Manenti}, \citenamefont {Manning}, \citenamefont
  {Marshall}, \citenamefont {Mohan}, \citenamefont {O\'Brien}, \citenamefont
  {Otterbach}, \citenamefont {Papageorge}, \citenamefont {Paquette},
  \citenamefont {Pelstring}, \citenamefont {Polloreno}, \citenamefont {Rawat},
  \citenamefont {Ryan}, \citenamefont {Renzas}, \citenamefont {Rubin},
  \citenamefont {Russell}, \citenamefont {Rust}, \citenamefont {Scarabelli},
  \citenamefont {Selvanayagam}, \citenamefont {Sinclair}, \citenamefont
  {Smith}, \citenamefont {Suska}, \citenamefont {To}, \citenamefont
  {Vahidpour}, \citenamefont {Vodrahalli}, \citenamefont {Whyland},
  \citenamefont {Yadav}, \citenamefont {Zeng},\ and\ \citenamefont
  {Rigetti}}]{Reagor2017}%
  \BibitemOpen
  \bibfield  {author} {\bibinfo {author} {\bibfnamefont {M.}~\bibnamefont
  {Reagor}}, \bibinfo {author} {\bibfnamefont {C.~B.}\ \bibnamefont {Osborn}},
  \bibinfo {author} {\bibfnamefont {N.}~\bibnamefont {Tezak}}, \bibinfo
  {author} {\bibfnamefont {A.}~\bibnamefont {Staley}}, \bibinfo {author}
  {\bibfnamefont {G.}~\bibnamefont {Prawiroatmodjo}}, \bibinfo {author}
  {\bibfnamefont {M.}~\bibnamefont {Scheer}}, \bibinfo {author} {\bibfnamefont
  {N.}~\bibnamefont {Alidoust}}, \bibinfo {author} {\bibfnamefont {E.~A.}\
  \bibnamefont {Sete}}, \bibinfo {author} {\bibfnamefont {N.}~\bibnamefont
  {Didier}}, \bibinfo {author} {\bibfnamefont {M.~P.}\ \bibnamefont
  {da~Silva}}, \bibinfo {author} {\bibfnamefont {E.}~\bibnamefont {Acala}},
  \bibinfo {author} {\bibfnamefont {J.}~\bibnamefont {Angeles}}, \bibinfo
  {author} {\bibfnamefont {A.}~\bibnamefont {Bestwick}}, \bibinfo {author}
  {\bibfnamefont {M.}~\bibnamefont {Block}}, \bibinfo {author} {\bibfnamefont
  {B.}~\bibnamefont {Bloom}}, \bibinfo {author} {\bibfnamefont
  {A.}~\bibnamefont {Bradley}}, \bibinfo {author} {\bibfnamefont
  {C.}~\bibnamefont {Bui}}, \bibinfo {author} {\bibfnamefont {S.}~\bibnamefont
  {Caldwell}}, \bibinfo {author} {\bibfnamefont {L.}~\bibnamefont {Capelluto}},
  \bibinfo {author} {\bibfnamefont {R.}~\bibnamefont {Chilcott}}, \bibinfo
  {author} {\bibfnamefont {J.}~\bibnamefont {Cordova}}, \bibinfo {author}
  {\bibfnamefont {G.}~\bibnamefont {Crossman}}, \bibinfo {author}
  {\bibfnamefont {M.}~\bibnamefont {Curtis}}, \bibinfo {author} {\bibfnamefont
  {S.}~\bibnamefont {Deshpande}}, \bibinfo {author} {\bibfnamefont {T.~E.}\
  \bibnamefont {Bouayadi}}, \bibinfo {author} {\bibfnamefont {D.}~\bibnamefont
  {Girshovich}}, \bibinfo {author} {\bibfnamefont {S.}~\bibnamefont {Hong}},
  \bibinfo {author} {\bibfnamefont {A.}~\bibnamefont {Hudson}}, \bibinfo
  {author} {\bibfnamefont {P.}~\bibnamefont {Karalekas}}, \bibinfo {author}
  {\bibfnamefont {K.}~\bibnamefont {Kuang}}, \bibinfo {author} {\bibfnamefont
  {M.}~\bibnamefont {Lenihan}}, \bibinfo {author} {\bibfnamefont
  {R.}~\bibnamefont {Manenti}}, \bibinfo {author} {\bibfnamefont
  {T.}~\bibnamefont {Manning}}, \bibinfo {author} {\bibfnamefont
  {J.}~\bibnamefont {Marshall}}, \bibinfo {author} {\bibfnamefont
  {Y.}~\bibnamefont {Mohan}}, \bibinfo {author} {\bibfnamefont
  {W.}~\bibnamefont {O\'Brien}}, \bibinfo {author} {\bibfnamefont
  {J.}~\bibnamefont {Otterbach}}, \bibinfo {author} {\bibfnamefont
  {A.}~\bibnamefont {Papageorge}}, \bibinfo {author} {\bibfnamefont {J.~.}\
  \bibnamefont {Paquette}}, \bibinfo {author} {\bibfnamefont {M.}~\bibnamefont
  {Pelstring}}, \bibinfo {author} {\bibfnamefont {A.}~\bibnamefont
  {Polloreno}}, \bibinfo {author} {\bibfnamefont {V.}~\bibnamefont {Rawat}},
  \bibinfo {author} {\bibfnamefont {C.~A.}\ \bibnamefont {Ryan}}, \bibinfo
  {author} {\bibfnamefont {R.}~\bibnamefont {Renzas}}, \bibinfo {author}
  {\bibfnamefont {N.}~\bibnamefont {Rubin}}, \bibinfo {author} {\bibfnamefont
  {D.}~\bibnamefont {Russell}}, \bibinfo {author} {\bibfnamefont
  {M.}~\bibnamefont {Rust}}, \bibinfo {author} {\bibfnamefont {D.}~\bibnamefont
  {Scarabelli}}, \bibinfo {author} {\bibfnamefont {M.}~\bibnamefont
  {Selvanayagam}}, \bibinfo {author} {\bibfnamefont {R.}~\bibnamefont
  {Sinclair}}, \bibinfo {author} {\bibfnamefont {R.}~\bibnamefont {Smith}},
  \bibinfo {author} {\bibfnamefont {M.}~\bibnamefont {Suska}}, \bibinfo
  {author} {\bibfnamefont {T.~.}\ \bibnamefont {To}}, \bibinfo {author}
  {\bibfnamefont {M.}~\bibnamefont {Vahidpour}}, \bibinfo {author}
  {\bibfnamefont {N.}~\bibnamefont {Vodrahalli}}, \bibinfo {author}
  {\bibfnamefont {T.}~\bibnamefont {Whyland}}, \bibinfo {author} {\bibfnamefont
  {K.}~\bibnamefont {Yadav}}, \bibinfo {author} {\bibfnamefont
  {W.}~\bibnamefont {Zeng}}, \ and\ \bibinfo {author} {\bibfnamefont {C.~T.}\
  \bibnamefont {Rigetti}},\ }\href {https://arxiv.org/abs/1706.06570}
  {\bibfield  {journal} {\bibinfo  {journal} {arXiv:1706.06570}\ } (\bibinfo
  {year} {2017})}\BibitemShut {NoStop}%
\bibitem [{\citenamefont {Bultink}\ \emph {et~al.}(2016)\citenamefont
  {Bultink}, \citenamefont {Rol}, \citenamefont {O'Brien}, \citenamefont {Fu},
  \citenamefont {Dikken}, \citenamefont {Dickel}, \citenamefont {Vermeulen},
  \citenamefont {de~Sterke}, \citenamefont {Bruno}, \citenamefont {Schouten},\
  and\ \citenamefont {DiCarlo}}]{Bultink2016}%
  \BibitemOpen
  \bibfield  {author} {\bibinfo {author} {\bibfnamefont {C.~C.}\ \bibnamefont
  {Bultink}}, \bibinfo {author} {\bibfnamefont {M.~A.}\ \bibnamefont {Rol}},
  \bibinfo {author} {\bibfnamefont {T.~E.}\ \bibnamefont {O'Brien}}, \bibinfo
  {author} {\bibfnamefont {X.}~\bibnamefont {Fu}}, \bibinfo {author}
  {\bibfnamefont {B.~C.~S.}\ \bibnamefont {Dikken}}, \bibinfo {author}
  {\bibfnamefont {C.}~\bibnamefont {Dickel}}, \bibinfo {author} {\bibfnamefont
  {R.~F.~L.}\ \bibnamefont {Vermeulen}}, \bibinfo {author} {\bibfnamefont
  {J.~C.}\ \bibnamefont {de~Sterke}}, \bibinfo {author} {\bibfnamefont
  {A.}~\bibnamefont {Bruno}}, \bibinfo {author} {\bibfnamefont {R.~N.}\
  \bibnamefont {Schouten}}, \ and\ \bibinfo {author} {\bibfnamefont
  {L.}~\bibnamefont {DiCarlo}},\ }\href {\doibase
  10.1103/PhysRevApplied.6.034008} {\bibfield  {journal} {\bibinfo  {journal}
  {Phys. Rev. Applied}\ }\textbf {\bibinfo {volume} {6}},\ \bibinfo {pages}
  {034008} (\bibinfo {year} {2016})}\BibitemShut {NoStop}%
\bibitem [{\citenamefont {Asaad}\ \emph {et~al.}(2016)\citenamefont {Asaad},
  \citenamefont {Dickel}, \citenamefont {Langford}, \citenamefont {Poletto},
  \citenamefont {Bruno}, \citenamefont {Rol}, \citenamefont {Deurloo},\ and\
  \citenamefont {DiCarlo}}]{Asaad2016}%
  \BibitemOpen
  \bibfield  {author} {\bibinfo {author} {\bibfnamefont {S.}~\bibnamefont
  {Asaad}}, \bibinfo {author} {\bibfnamefont {C.}~\bibnamefont {Dickel}},
  \bibinfo {author} {\bibfnamefont {N.~K.}\ \bibnamefont {Langford}}, \bibinfo
  {author} {\bibfnamefont {S.}~\bibnamefont {Poletto}}, \bibinfo {author}
  {\bibfnamefont {A.}~\bibnamefont {Bruno}}, \bibinfo {author} {\bibfnamefont
  {M.~A.}\ \bibnamefont {Rol}}, \bibinfo {author} {\bibfnamefont
  {D.}~\bibnamefont {Deurloo}}, \ and\ \bibinfo {author} {\bibfnamefont
  {L.}~\bibnamefont {DiCarlo}},\ }\href
  {http://dx.doi.org/10.1038/npjqi.2016.29} {\bibfield  {journal} {\bibinfo
  {journal} {Npj Quantum Information}\ }\textbf {\bibinfo {volume} {2}},\
  \bibinfo {pages} {16029} (\bibinfo {year} {2016})}\BibitemShut {NoStop}%
\bibitem [{\citenamefont {Song}\ \emph {et~al.}(2017)\citenamefont {Song},
  \citenamefont {Xu}, \citenamefont {Liu}, \citenamefont {Yang}, \citenamefont
  {Zheng}, \citenamefont {Deng}, \citenamefont {Xie}, \citenamefont {Huang},
  \citenamefont {Guo}, \citenamefont {Zhang}, \citenamefont {Zhang},
  \citenamefont {Xu}, \citenamefont {Zheng}, \citenamefont {Zhu}, \citenamefont
  {Wang}, \citenamefont {Chen}, \citenamefont {Lu}, \citenamefont {Han},\ and\
  \citenamefont {Pan}}]{Song2017}%
  \BibitemOpen
  \bibfield  {author} {\bibinfo {author} {\bibfnamefont {C.}~\bibnamefont
  {Song}}, \bibinfo {author} {\bibfnamefont {K.}~\bibnamefont {Xu}}, \bibinfo
  {author} {\bibfnamefont {W.}~\bibnamefont {Liu}}, \bibinfo {author}
  {\bibfnamefont {C.-p.}\ \bibnamefont {Yang}}, \bibinfo {author}
  {\bibfnamefont {S.-B.}\ \bibnamefont {Zheng}}, \bibinfo {author}
  {\bibfnamefont {H.}~\bibnamefont {Deng}}, \bibinfo {author} {\bibfnamefont
  {Q.}~\bibnamefont {Xie}}, \bibinfo {author} {\bibfnamefont {K.}~\bibnamefont
  {Huang}}, \bibinfo {author} {\bibfnamefont {Q.}~\bibnamefont {Guo}}, \bibinfo
  {author} {\bibfnamefont {L.}~\bibnamefont {Zhang}}, \bibinfo {author}
  {\bibfnamefont {P.}~\bibnamefont {Zhang}}, \bibinfo {author} {\bibfnamefont
  {D.}~\bibnamefont {Xu}}, \bibinfo {author} {\bibfnamefont {D.}~\bibnamefont
  {Zheng}}, \bibinfo {author} {\bibfnamefont {X.}~\bibnamefont {Zhu}}, \bibinfo
  {author} {\bibfnamefont {H.}~\bibnamefont {Wang}}, \bibinfo {author}
  {\bibfnamefont {Y.-A.}\ \bibnamefont {Chen}}, \bibinfo {author}
  {\bibfnamefont {C.-Y.}\ \bibnamefont {Lu}}, \bibinfo {author} {\bibfnamefont
  {S.}~\bibnamefont {Han}}, \ and\ \bibinfo {author} {\bibfnamefont {J.-W.}\
  \bibnamefont {Pan}},\ }\href {\doibase 10.1103/PhysRevLett.119.180511}
  {\bibfield  {journal} {\bibinfo  {journal} {Phys. Rev. Lett.}\ }\textbf
  {\bibinfo {volume} {119}},\ \bibinfo {pages} {180511} (\bibinfo {year}
  {2017})}\BibitemShut {NoStop}%
\bibitem [{\citenamefont {Fowler}\ \emph {et~al.}(2012)\citenamefont {Fowler},
  \citenamefont {Mariantoni}, \citenamefont {Martinis},\ and\ \citenamefont
  {Cleland}}]{Fowler2012}%
  \BibitemOpen
  \bibfield  {author} {\bibinfo {author} {\bibfnamefont {A.~G.}\ \bibnamefont
  {Fowler}}, \bibinfo {author} {\bibfnamefont {M.}~\bibnamefont {Mariantoni}},
  \bibinfo {author} {\bibfnamefont {J.~M.}\ \bibnamefont {Martinis}}, \ and\
  \bibinfo {author} {\bibfnamefont {A.~N.}\ \bibnamefont {Cleland}},\ }\href
  {\doibase 10.1103/PhysRevA.86.032324} {\bibfield  {journal} {\bibinfo
  {journal} {Phys. Rev. A}\ }\textbf {\bibinfo {volume} {86}},\ \bibinfo
  {pages} {032324} (\bibinfo {year} {2012})}\BibitemShut {NoStop}%
\bibitem [{\citenamefont {Griffiths}\ and\ \citenamefont
  {Niu}(1996)}]{Griffiths1996}%
  \BibitemOpen
  \bibfield  {author} {\bibinfo {author} {\bibfnamefont {R.~B.}\ \bibnamefont
  {Griffiths}}\ and\ \bibinfo {author} {\bibfnamefont {C.-S.}\ \bibnamefont
  {Niu}},\ }\href {\doibase 10.1103/PhysRevLett.76.3228} {\bibfield  {journal}
  {\bibinfo  {journal} {Phys. Rev. Lett.}\ }\textbf {\bibinfo {volume} {76}},\
  \bibinfo {pages} {3228} (\bibinfo {year} {1996})}\BibitemShut {NoStop}%
\bibitem [{\citenamefont {Bennett}\ \emph {et~al.}(1996)\citenamefont
  {Bennett}, \citenamefont {Brassard}, \citenamefont {Popescu}, \citenamefont
  {Schumacher}, \citenamefont {Smolin},\ and\ \citenamefont
  {Wootters}}]{Bennett1996b}%
  \BibitemOpen
  \bibfield  {author} {\bibinfo {author} {\bibfnamefont {C.~H.}\ \bibnamefont
  {Bennett}}, \bibinfo {author} {\bibfnamefont {G.}~\bibnamefont {Brassard}},
  \bibinfo {author} {\bibfnamefont {S.}~\bibnamefont {Popescu}}, \bibinfo
  {author} {\bibfnamefont {B.}~\bibnamefont {Schumacher}}, \bibinfo {author}
  {\bibfnamefont {J.~A.}\ \bibnamefont {Smolin}}, \ and\ \bibinfo {author}
  {\bibfnamefont {W.~K.}\ \bibnamefont {Wootters}},\ }\href {\doibase
  10.1103/PhysRevLett.76.722} {\bibfield  {journal} {\bibinfo  {journal} {Phys.
  Rev. Lett.}\ }\textbf {\bibinfo {volume} {76}},\ \bibinfo {pages} {722}
  (\bibinfo {year} {1996})}\BibitemShut {NoStop}%
\bibitem [{\citenamefont {Yurke}\ and\ \citenamefont
  {Stoler}(1992)}]{Yurke1992}%
  \BibitemOpen
  \bibfield  {author} {\bibinfo {author} {\bibfnamefont {B.}~\bibnamefont
  {Yurke}}\ and\ \bibinfo {author} {\bibfnamefont {D.}~\bibnamefont {Stoler}},\
  }\href {\doibase 10.1103/PhysRevA.46.2229} {\bibfield  {journal} {\bibinfo
  {journal} {Phys. Rev. A}\ }\textbf {\bibinfo {volume} {46}},\ \bibinfo
  {pages} {2229} (\bibinfo {year} {1992})}\BibitemShut {NoStop}%
\bibitem [{\citenamefont {Sete}\ \emph {et~al.}(2015)\citenamefont {Sete},
  \citenamefont {Martinis},\ and\ \citenamefont {Korotkov}}]{Sete2015}%
  \BibitemOpen
  \bibfield  {author} {\bibinfo {author} {\bibfnamefont {E.~A.}\ \bibnamefont
  {Sete}}, \bibinfo {author} {\bibfnamefont {J.~M.}\ \bibnamefont {Martinis}},
  \ and\ \bibinfo {author} {\bibfnamefont {A.~N.}\ \bibnamefont {Korotkov}},\
  }\href {\doibase 10.1103/PhysRevA.92.012325} {\bibfield  {journal} {\bibinfo
  {journal} {Phys. Rev. A}\ }\textbf {\bibinfo {volume} {92}},\ \bibinfo
  {pages} {012325} (\bibinfo {year} {2015})}\BibitemShut {NoStop}%
\bibitem [{\citenamefont {Gambetta}\ \emph {et~al.}(2006)\citenamefont
  {Gambetta}, \citenamefont {Blais}, \citenamefont {Schuster}, \citenamefont
  {Wallraff}, \citenamefont {Frunzio}, \citenamefont {Majer}, \citenamefont
  {Devoret}, \citenamefont {Girvin},\ and\ \citenamefont
  {Schoelkopf}}]{Gambetta2006}%
  \BibitemOpen
  \bibfield  {author} {\bibinfo {author} {\bibfnamefont {J.}~\bibnamefont
  {Gambetta}}, \bibinfo {author} {\bibfnamefont {A.}~\bibnamefont {Blais}},
  \bibinfo {author} {\bibfnamefont {D.~I.}\ \bibnamefont {Schuster}}, \bibinfo
  {author} {\bibfnamefont {A.}~\bibnamefont {Wallraff}}, \bibinfo {author}
  {\bibfnamefont {L.}~\bibnamefont {Frunzio}}, \bibinfo {author} {\bibfnamefont
  {J.}~\bibnamefont {Majer}}, \bibinfo {author} {\bibfnamefont {M.~H.}\
  \bibnamefont {Devoret}}, \bibinfo {author} {\bibfnamefont {S.~M.}\
  \bibnamefont {Girvin}}, \ and\ \bibinfo {author} {\bibfnamefont {R.~J.}\
  \bibnamefont {Schoelkopf}},\ }\href {\doibase 10.1103/PhysRevA.74.042318}
  {\bibfield  {journal} {\bibinfo  {journal} {Phys. Rev. A}\ }\textbf {\bibinfo
  {volume} {74}},\ \bibinfo {pages} {042318} (\bibinfo {year}
  {2006})}\BibitemShut {NoStop}%
\bibitem [{\citenamefont {Koch}\ \emph {et~al.}(2007)\citenamefont {Koch},
  \citenamefont {Yu}, \citenamefont {Gambetta}, \citenamefont {Houck},
  \citenamefont {Schuster}, \citenamefont {Majer}, \citenamefont {Blais},
  \citenamefont {Devoret}, \citenamefont {Girvin},\ and\ \citenamefont
  {Schoelkopf}}]{Koch2007}%
  \BibitemOpen
  \bibfield  {author} {\bibinfo {author} {\bibfnamefont {J.}~\bibnamefont
  {Koch}}, \bibinfo {author} {\bibfnamefont {T.~M.}\ \bibnamefont {Yu}},
  \bibinfo {author} {\bibfnamefont {J.}~\bibnamefont {Gambetta}}, \bibinfo
  {author} {\bibfnamefont {A.~A.}\ \bibnamefont {Houck}}, \bibinfo {author}
  {\bibfnamefont {D.~I.}\ \bibnamefont {Schuster}}, \bibinfo {author}
  {\bibfnamefont {J.}~\bibnamefont {Majer}}, \bibinfo {author} {\bibfnamefont
  {A.}~\bibnamefont {Blais}}, \bibinfo {author} {\bibfnamefont {M.~H.}\
  \bibnamefont {Devoret}}, \bibinfo {author} {\bibfnamefont {S.~M.}\
  \bibnamefont {Girvin}}, \ and\ \bibinfo {author} {\bibfnamefont {R.~J.}\
  \bibnamefont {Schoelkopf}},\ }\href {\doibase 10.1103/PhysRevA.76.042319}
  {\bibfield  {journal} {\bibinfo  {journal} {Phys. Rev. A}\ }\textbf {\bibinfo
  {volume} {76}},\ \bibinfo {eid} {042319} (\bibinfo {year}
  {2007})}\BibitemShut {NoStop}%
\bibitem [{\citenamefont {Barends}\ \emph {et~al.}(2013)\citenamefont
  {Barends}, \citenamefont {Kelly}, \citenamefont {Megrant}, \citenamefont
  {Sank}, \citenamefont {Jeffrey}, \citenamefont {Chen}, \citenamefont {Yin},
  \citenamefont {Chiaro}, \citenamefont {Mutus}, \citenamefont {Neill},
  \citenamefont {O'Malley}, \citenamefont {Roushan}, \citenamefont {Wenner},
  \citenamefont {White}, \citenamefont {Cleland},\ and\ \citenamefont
  {Martinis}}]{Barends2013}%
  \BibitemOpen
  \bibfield  {author} {\bibinfo {author} {\bibfnamefont {R.}~\bibnamefont
  {Barends}}, \bibinfo {author} {\bibfnamefont {J.}~\bibnamefont {Kelly}},
  \bibinfo {author} {\bibfnamefont {A.}~\bibnamefont {Megrant}}, \bibinfo
  {author} {\bibfnamefont {D.}~\bibnamefont {Sank}}, \bibinfo {author}
  {\bibfnamefont {E.}~\bibnamefont {Jeffrey}}, \bibinfo {author} {\bibfnamefont
  {Y.}~\bibnamefont {Chen}}, \bibinfo {author} {\bibfnamefont {Y.}~\bibnamefont
  {Yin}}, \bibinfo {author} {\bibfnamefont {B.}~\bibnamefont {Chiaro}},
  \bibinfo {author} {\bibfnamefont {J.}~\bibnamefont {Mutus}}, \bibinfo
  {author} {\bibfnamefont {C.}~\bibnamefont {Neill}}, \bibinfo {author}
  {\bibfnamefont {P.}~\bibnamefont {O'Malley}}, \bibinfo {author}
  {\bibfnamefont {P.}~\bibnamefont {Roushan}}, \bibinfo {author} {\bibfnamefont
  {J.}~\bibnamefont {Wenner}}, \bibinfo {author} {\bibfnamefont {T.~C.}\
  \bibnamefont {White}}, \bibinfo {author} {\bibfnamefont {A.~N.}\ \bibnamefont
  {Cleland}}, \ and\ \bibinfo {author} {\bibfnamefont {J.~M.}\ \bibnamefont
  {Martinis}},\ }\href {http://link.aps.org/doi/10.1103/PhysRevLett.111.080502}
  {\bibfield  {journal} {\bibinfo  {journal} {Phys. Rev. Lett.}\ }\textbf
  {\bibinfo {volume} {111}},\ \bibinfo {pages} {080502} (\bibinfo {year}
  {2013})}\BibitemShut {NoStop}%
\bibitem [{\citenamefont {Gambetta}\ \emph {et~al.}(2007)\citenamefont
  {Gambetta}, \citenamefont {Braff}, \citenamefont {Wallraff}, \citenamefont
  {Girvin},\ and\ \citenamefont {Schoelkopf}}]{Gambetta2007}%
  \BibitemOpen
  \bibfield  {author} {\bibinfo {author} {\bibfnamefont {J.}~\bibnamefont
  {Gambetta}}, \bibinfo {author} {\bibfnamefont {W.~A.}\ \bibnamefont {Braff}},
  \bibinfo {author} {\bibfnamefont {A.}~\bibnamefont {Wallraff}}, \bibinfo
  {author} {\bibfnamefont {S.~M.}\ \bibnamefont {Girvin}}, \ and\ \bibinfo
  {author} {\bibfnamefont {R.~J.}\ \bibnamefont {Schoelkopf}},\ }\href
  {\doibase 10.1103/PhysRevA.76.012325} {\bibfield  {journal} {\bibinfo
  {journal} {Phys. Rev. A}\ }\textbf {\bibinfo {volume} {76}},\ \bibinfo
  {pages} {012325} (\bibinfo {year} {2007})}\BibitemShut {NoStop}%
\bibitem [{\citenamefont {Bultink}\ \emph {et~al.}(2017)\citenamefont
  {Bultink}, \citenamefont {Tarasinski}, \citenamefont {Haandbaek},
  \citenamefont {Poletto}, \citenamefont {Haider}, \citenamefont {Michalak},
  \citenamefont {Bruno},\ and\ \citenamefont {DiCarlo}}]{Bultink2017}%
  \BibitemOpen
  \bibfield  {author} {\bibinfo {author} {\bibfnamefont {C.~C.}\ \bibnamefont
  {Bultink}}, \bibinfo {author} {\bibfnamefont {B.}~\bibnamefont {Tarasinski}},
  \bibinfo {author} {\bibfnamefont {N.}~\bibnamefont {Haandbaek}}, \bibinfo
  {author} {\bibfnamefont {S.}~\bibnamefont {Poletto}}, \bibinfo {author}
  {\bibfnamefont {N.}~\bibnamefont {Haider}}, \bibinfo {author} {\bibfnamefont
  {D.~J.}\ \bibnamefont {Michalak}}, \bibinfo {author} {\bibfnamefont
  {A.}~\bibnamefont {Bruno}}, \ and\ \bibinfo {author} {\bibfnamefont
  {L.}~\bibnamefont {DiCarlo}},\ }\href {https://arxiv.org/abs/1711.05336}
  {\bibfield  {journal} {\bibinfo  {journal} {arXiv:1711.05336}\ } (\bibinfo
  {year} {2017})}\BibitemShut {NoStop}%
\bibitem [{\citenamefont {Jin}\ \emph {et~al.}(2015)\citenamefont {Jin},
  \citenamefont {Kamal}, \citenamefont {Sears}, \citenamefont {Gudmundsen},
  \citenamefont {Hover}, \citenamefont {Miloshi}, \citenamefont {Slattery},
  \citenamefont {Yan}, \citenamefont {Yoder}, \citenamefont {Orlando},
  \citenamefont {Gustavsson},\ and\ \citenamefont {Oliver}}]{Jin2015b}%
  \BibitemOpen
  \bibfield  {author} {\bibinfo {author} {\bibfnamefont {X.~Y.}\ \bibnamefont
  {Jin}}, \bibinfo {author} {\bibfnamefont {A.}~\bibnamefont {Kamal}}, \bibinfo
  {author} {\bibfnamefont {A.~P.}\ \bibnamefont {Sears}}, \bibinfo {author}
  {\bibfnamefont {T.}~\bibnamefont {Gudmundsen}}, \bibinfo {author}
  {\bibfnamefont {D.}~\bibnamefont {Hover}}, \bibinfo {author} {\bibfnamefont
  {J.}~\bibnamefont {Miloshi}}, \bibinfo {author} {\bibfnamefont
  {R.}~\bibnamefont {Slattery}}, \bibinfo {author} {\bibfnamefont
  {F.}~\bibnamefont {Yan}}, \bibinfo {author} {\bibfnamefont {J.}~\bibnamefont
  {Yoder}}, \bibinfo {author} {\bibfnamefont {T.~P.}\ \bibnamefont {Orlando}},
  \bibinfo {author} {\bibfnamefont {S.}~\bibnamefont {Gustavsson}}, \ and\
  \bibinfo {author} {\bibfnamefont {W.~D.}\ \bibnamefont {Oliver}},\ }\href
  {\doibase http://dx.doi.org/10.1103/PhysRevLett.114.240501} {\bibfield
  {journal} {\bibinfo  {journal} {Phys. Rev. Lett.}\ }\textbf {\bibinfo
  {volume} {114}},\ \bibinfo {pages} {240501} (\bibinfo {year}
  {2015})}\BibitemShut {NoStop}%
\bibitem [{\citenamefont {Boissonneault}\ \emph {et~al.}(2009)\citenamefont
  {Boissonneault}, \citenamefont {Gambetta},\ and\ \citenamefont
  {Blais}}]{Boissonneault2009}%
  \BibitemOpen
  \bibfield  {author} {\bibinfo {author} {\bibfnamefont {M.}~\bibnamefont
  {Boissonneault}}, \bibinfo {author} {\bibfnamefont {J.~M.}\ \bibnamefont
  {Gambetta}}, \ and\ \bibinfo {author} {\bibfnamefont {A.}~\bibnamefont
  {Blais}},\ }\href {\doibase 10.1103/PhysRevA.79.013819} {\bibfield  {journal}
  {\bibinfo  {journal} {Phys. Rev. A}\ }\textbf {\bibinfo {volume} {79}},\
  \bibinfo {eid} {013819} (\bibinfo {year} {2009})}\BibitemShut {NoStop}%
\bibitem [{\citenamefont {Slichter}\ \emph {et~al.}(2012)\citenamefont
  {Slichter}, \citenamefont {Vijay}, \citenamefont {Weber}, \citenamefont
  {Boutin}, \citenamefont {Boissonneault}, \citenamefont {Gambetta},
  \citenamefont {Blais},\ and\ \citenamefont {Siddiqi}}]{Slichter2012}%
  \BibitemOpen
  \bibfield  {author} {\bibinfo {author} {\bibfnamefont {D.~H.}\ \bibnamefont
  {Slichter}}, \bibinfo {author} {\bibfnamefont {R.}~\bibnamefont {Vijay}},
  \bibinfo {author} {\bibfnamefont {S.~J.}\ \bibnamefont {Weber}}, \bibinfo
  {author} {\bibfnamefont {S.}~\bibnamefont {Boutin}}, \bibinfo {author}
  {\bibfnamefont {M.}~\bibnamefont {Boissonneault}}, \bibinfo {author}
  {\bibfnamefont {J.~M.}\ \bibnamefont {Gambetta}}, \bibinfo {author}
  {\bibfnamefont {A.}~\bibnamefont {Blais}}, \ and\ \bibinfo {author}
  {\bibfnamefont {I.}~\bibnamefont {Siddiqi}},\ }\href {\doibase
  10.1103/PhysRevLett.109.153601} {\bibfield  {journal} {\bibinfo  {journal}
  {Phys. Rev. Lett.}\ }\textbf {\bibinfo {volume} {109}},\ \bibinfo {pages}
  {153601} (\bibinfo {year} {2012})}\BibitemShut {NoStop}%
\bibitem [{\citenamefont {Versluis}\ \emph {et~al.}(2017)\citenamefont
  {Versluis}, \citenamefont {Poletto}, \citenamefont {Khammassi}, \citenamefont
  {Tarasinski}, \citenamefont {Haider}, \citenamefont {Michalak}, \citenamefont
  {Bruno}, \citenamefont {Bertels},\ and\ \citenamefont
  {DiCarlo}}]{Versluis2017}%
  \BibitemOpen
  \bibfield  {author} {\bibinfo {author} {\bibfnamefont {R.}~\bibnamefont
  {Versluis}}, \bibinfo {author} {\bibfnamefont {S.}~\bibnamefont {Poletto}},
  \bibinfo {author} {\bibfnamefont {N.}~\bibnamefont {Khammassi}}, \bibinfo
  {author} {\bibfnamefont {B.}~\bibnamefont {Tarasinski}}, \bibinfo {author}
  {\bibfnamefont {N.}~\bibnamefont {Haider}}, \bibinfo {author} {\bibfnamefont
  {D.~J.}\ \bibnamefont {Michalak}}, \bibinfo {author} {\bibfnamefont
  {A.}~\bibnamefont {Bruno}}, \bibinfo {author} {\bibfnamefont
  {K.}~\bibnamefont {Bertels}}, \ and\ \bibinfo {author} {\bibfnamefont
  {L.}~\bibnamefont {DiCarlo}},\ }\href {\doibase
  10.1103/PhysRevApplied.8.034021} {\bibfield  {journal} {\bibinfo  {journal}
  {Phys. Rev. Applied}\ }\textbf {\bibinfo {volume} {8}},\ \bibinfo {pages}
  {034021} (\bibinfo {year} {2017})}\BibitemShut {NoStop}%
\bibitem [{\citenamefont {Gardiner}\ and\ \citenamefont
  {Collett}(1985)}]{Gardiner1985}%
  \BibitemOpen
  \bibfield  {author} {\bibinfo {author} {\bibfnamefont {C.~W.}\ \bibnamefont
  {Gardiner}}\ and\ \bibinfo {author} {\bibfnamefont {M.~J.}\ \bibnamefont
  {Collett}},\ }\href {\doibase 10.1103/PhysRevA.31.3761} {\bibfield  {journal}
  {\bibinfo  {journal} {Phys. Rev. A}\ }\textbf {\bibinfo {volume} {31}},\
  \bibinfo {pages} {3761} (\bibinfo {year} {1985})}\BibitemShut {NoStop}%
\bibitem [{\citenamefont {Bourassa}\ \emph {et~al.}(2012)\citenamefont
  {Bourassa}, \citenamefont {Beaudoin}, \citenamefont {Gambetta},\ and\
  \citenamefont {Blais}}]{Bourassa2012}%
  \BibitemOpen
  \bibfield  {author} {\bibinfo {author} {\bibfnamefont {J.}~\bibnamefont
  {Bourassa}}, \bibinfo {author} {\bibfnamefont {F.}~\bibnamefont {Beaudoin}},
  \bibinfo {author} {\bibfnamefont {J.~M.}\ \bibnamefont {Gambetta}}, \ and\
  \bibinfo {author} {\bibfnamefont {A.}~\bibnamefont {Blais}},\ }\href
  {\doibase 10.1103/PhysRevA.86.013814} {\bibfield  {journal} {\bibinfo
  {journal} {Phys. Rev. A}\ }\textbf {\bibinfo {volume} {86}},\ \bibinfo
  {pages} {013814} (\bibinfo {year} {2012})}\BibitemShut {NoStop}%
\end{thebibliography}%

\end{document}